\documentclass[aps,prl,reprint,nobibnotes,superscriptaddress]{revtex4-2}
\usepackage{amsmath,amssymb,amsfonts}
\usepackage{graphicx}
\usepackage{bm}           % bold math
\usepackage{xcolor}
\usepackage{hyperref}
\usepackage{cleveref}
\usepackage{physics}      % \ket, \bra, \ev, \mel, etc.  (remove if unused)
\usepackage{dcolumn}      % align table columns on decimal point

\hypersetup{
  colorlinks = true,
  linkcolor  = blue,
  citecolor  = blue,
  urlcolor   = blue,
}
\usepackage{xcolor}
\definecolor{green2}{rgb}{0., 0.5, 0.}
\definecolor{blue2}{rgb}{0., 0.35, 1}

\usepackage{amsmath,amssymb}
\usepackage{graphicx}
\usepackage{float}
\usepackage[switch]{lineno}
\usepackage{braket}
\usepackage{hyperref}

\hypersetup{
  colorlinks = true,
  linkcolor  = blue,
  citecolor  = blue,
  urlcolor = blue,
}

\newcommand{\vtg}{V_\mathrm{{TG}}}
\newcommand{\vbg}{V_\mathrm{{BG}}}
\newcommand{\bfA}{{\textbf{a}}}
\newcommand{\bfB}{{\textbf{b}}}
\newcommand{\bfC}{{\textbf{c}}}
\newcommand{\bfD}{{\textbf{d}}}
\newcommand{\bfE}{{\textbf{e}}}
\newcommand{\bfF}{{\textbf{f}}}

\newcommand{\W}{{WSe\(_2\)}}
\newcommand{\Mo}{{MoSe\(_2\)}}

\newcommand{\varn}{{\text{Var}(N)}}
\newcommand{\figone}{{Fig.\,\ref{fig:one}}}
\newcommand{\figtwo}{{Fig.\,\ref{fig:two}}}
\newcommand{\figthree}{{Fig.\,\ref{fig:three}}}
\newcommand{\figfour}{{Fig.\,\ref{fig:four}}}

\begin{document}

\title{\Large Photon correlation microscopy of quantum matter}

\author{Elie Vandoolaeghe}
\thanks{These authors contributed equally}
\affiliation{Institute for Quantum Electronics, ETH Z\"urich, CH-8093 Z\"urich, Switzerland}

\author{I\~{n}igo Lasheras}
\thanks{These authors contributed equally}
\affiliation{Institute for Quantum Electronics, ETH Z\"urich, CH-8093 Z\"urich, Switzerland}

\author{Chirag Vaswani}
\affiliation{NTT Research, Inc. Physics \& Informatics Laboratories, 940 Stewart Dr, Sunnyvale, CA 94085}

\author{Sampriti Saha}
\affiliation{Institute for Theoretical Physics, ETH Z\"urich, CH-8093 Z\"urich, Switzerland}

\author{Purbasha Ray}
\affiliation{Quantum Materials and Device Research Lab, Materials Research Center, Indian Institute of Technology, Kharagpur, India}

\author{Takashi Taniguchi}
\affiliation{National Institute for Materials Science, Namiki 1-1, Tsukuba, 305-0044, Ibaraki, Japan}

\author{Kenji Watanabe}
\affiliation{National Institute for Materials Science, Namiki 1-1, Tsukuba, 305-0044, Ibaraki, Japan}

\author{Prasana Sahoo}
\affiliation{Quantum Materials and Device Research Lab, Materials Research Center, Indian Institute of Technology, Kharagpur, India}

\author{Nicol\`o Defenu}
\affiliation{Institute for Theoretical Physics, ETH Z\"urich, CH-8093 Z\"urich, Switzerland}

\author{Thibault Chervy}
\email{thibault.chervy@ntt-research.com}
\affiliation{NTT Research, Inc. Physics \& Informatics Laboratories, 940 Stewart Dr, Sunnyvale, CA 94085}

\author{Puneet A. Murthy}
\email{murthyp@ethz.ch}
\affiliation{Institute for Quantum Electronics, ETH Z\"urich, CH-8093 Z\"urich, Switzerland}

\maketitle
\textbf{Light and matter share fundamental statistical properties, yet the experimental probes of quantum optics and many-body physics have largely evolved along separate trajectories. While many-body physics explores emergent collective phenomena, quantum optics has refined the measurement of correlations between individual photons. Here, we introduce photon correlation microscopy (PCM) -- which bridges the two domains by leveraging correlations of emitted light to probe the correlations in quantum matter at mesoscopic scales. We demonstrate this approach using a one-dimensional (1D) ensemble of dipolar excitons confined at a lateral monolayer MoSe$_2$-WSe$_2$ heterojunction. We use gate-defined potentials to confine the 1D excitons to a mesoscopic lengthscale to enhance the visibility of matter correlations in the emitted photon field. Power-dependent spectroscopy reveals a transition from a compressible to an incompressible phase, signaled by the simultaneous saturation of the emission intensity and energy blueshift, which is supported by numerical simulations. Through this crossover, photon correlation measurements show a striking evolution from bunching at low densities to antibunching at high densities. This constitutes a many-body blockade of photon emission emerging directly from a number-stabilized state, driven by collective dipolar repulsion.  Our results establish PCM as a powerful probe of many-body physics through the lens of quantum optics, extensible to a broad class of correlated electronic phases, while pointing toward a route to generating non-classical light through many-body correlations.}

Correlations lie at the heart of both many-body physics and quantum optics. In condensed matter, the pair correlation function $g^{(2)}_\text{mat}(r)$ and the number variance characterize the spatial and statistical organization of a many-body state, distinguishing a Fermi liquid from a Wigner crystal, a superfluid from a Mott insulator, a Luttinger liquid from a thermal gas \cite{Bruus2002}. Photon correlations play an analogously foundational role in quantum optics.
Bunched, coherent, and antibunched statistics — measuring whether photons arrive in clusters, randomly, or one at a time — distinguish thermal, laser, and single-photon sources \cite{Mandel1995}, and Hanbury Brown--Twiss (HBT) measurements of the second-order optical coherence $g^{(2)}_\text{ph}(\tau)$ constitutes a standard diagnostic of non-classical light.

Despite this shared centrality, the two fields have developed largely independent experimental toolkits. Many-body matter is characterized through a rich array of transport, scattering, and spectroscopic probes, while direct real-space access to higher-order matter correlations remains experimentally challenging --- notable exceptions include electron shot-noise spectroscopy in mesoscopic conductors, which resolves current fluctuations to access charge correlations \cite{Blanter2001} and atomic quantum gas experiments, which image local density distributions of particles \cite{Bloch2008, Folling2005, Bakr2009, Hilker2017, Bergschneider2019, Holten2022, Yao2025}. Quantum optics, on the other hand, has developed exquisite tools for generating non-classical states of light and resolving their temporal correlations. 
However, the standard paradigm for generating non-classical light has largely fallen into two categories \cite{Chang2014}: isolated single emitters such as atoms, quantum dots, and color centers \cite{Kimble1977, Hennessy2007, Aharonovich2016, Toninelli2021}, or interaction-induced blockade in resonantly driven ensembles, such as Rydberg-EIT media \cite{Peyronel2012, Weber2015} and exciton-polariton systems \cite{Delteil2019, MunozMatutano2019}. 

\begin{figure*}[!t]
\centering
\includegraphics[]{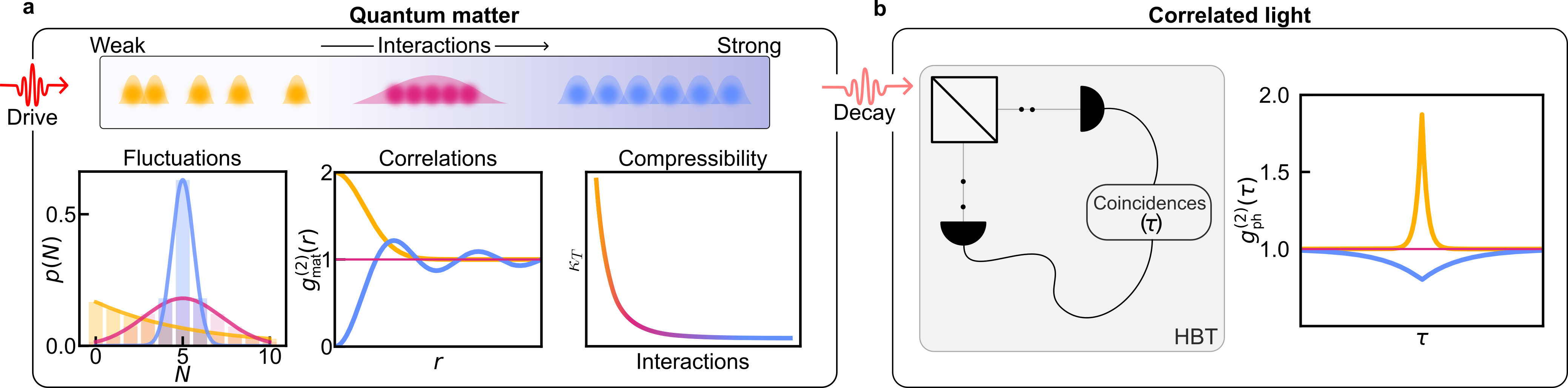}

\caption{\textbf{Photon correlation microscopy of many-body systems.}
(\bfA) \emph{Matter.} A mesoscopic many-body system of emitters exhibiting three distinct phases: a compressible thermal phase (yellow), a quantum-degenerate phase (purple), and a strongly correlated, incompressible phase (blue). These are mirrored in three complementary observables (left to right): the number distribution $p(N)$ broadens in the thermal phase and sharpens toward the incompressible regime; the matter pair correlator $g^{(2)}_\text{mat}(r)$ evolves (for bosons) from spatial bunching to a deepening correlation hole; the compressibility $\kappa_T$ collapses as repulsive interactions stiffen the system.
(\bfB) \emph{Light.} Emitted photons are sent into a Hanbury Brown--Twiss interferometer. The thermal regime yields bunching ($g^{(2)}_\text{ph}(0) > 1$), the quantum-degenerate regime Poissonian statistics ($g^{(2)}_\text{ph}(0) \approx 1$), and the incompressible regime antibunching ($g^{(2)}_\text{ph}(0) < 1$). The temporal photon correlation thus serves as a direct optical readout of the many-body correlations of the underlying matter --- as per Eq.~\ref{eq:g2}.
}
\label{fig:one} 
\end{figure*}

Here, we introduce \emph{Photon Correlation Microscopy}, which bridges the two approaches, where non-classical photon statistics emerge from many-body correlations in quantum matter, and hence can be used to directly probe them. The link between photon and matter correlations rests on a general thermodynamic principle, illustrated in \figone. As a many-body system (\figone\,\bfA) is driven from a thermal phase toward a strongly correlated regime, the energy cost of adding a particle freezes it into an increasingly stiff configuration. This emergent rigidity manifests both in particle number --- suppressed density fluctuations --- and in space, where a correlation hole develops at short distances. In equilibrium, the isothermal compressibility $\kappa_T$ at temperature $T$, number variance ($\Delta N^2$), and spatial pair correlator $g^{(2)}_\text{mat}(x,x')$ are therefore equivalent observables, linked exactly through the fluctuation–dissipation theorem \cite{Pathria2021}; measuring any one determines the other two.

When the constituents of such a system are quantum emitters (atoms, molecules, excitons) whose photon emission rate is proportional to the particle number $N$, the temporal statistics of the emitted photons directly inherit the number fluctuations of the matter \cite{Teich1988,Mandel1995} (\figone\,\bfB). 
The central concept underlying PCM is that the normalized zero-delay photon correlation function, $g^{(2)}_\text{ph}(\tau = 0)$, provides unified access to all three matter observables — fluctuations, spatial correlations, and compressibility — in a single measurement: 
\begin{align}
g^{(2)}_\text{ph}(0)\, = \,
& \underbrace{\frac{\langle N(N-1)\rangle}{\langle N\rangle^{2}}}_{\text{Fluctuations}} \nonumber\\
=\;& \underbrace{\frac{1}{L^2} \!\iint_{L}\! g^{(2)}_\text{mat}(x,x')\,dx\,dx'}_{\text{Correlations}} \nonumber\\
\overset{eq.}{=}\;& \underbrace{1\;+\;\frac{k_{B}T\kappa_{T}}{L}\;-\;\frac{1}{\langle N\rangle}}_{\text{Compressibility}},
\label{eq:g2}
\end{align}

where the last identity holds in thermal equilibrium. A remarkable outcome of Eq.\,\ref{eq:g2} is that the spatial density-density correlation function of matter --- a four-point correlator ---  becomes accessible non-invasively, averaged over a window $L$, through the photon correlation signal. Two experimental parameters, the average particle number $\braket{N}$ and the sampling length $L$, then determine the detectable contrast ($g^{(2)}_\text{ph}(0) - 1$) of the PCM. For bosons, a compressible thermal phase exhibits enhanced spatial correlations (bunching) over a coherence length $\xi_\text{coh}$, while a strongly correlated phase develops a \emph{correlation hole} in $g^{(2)}_\text{mat}(x,x')$ for repulsive interactions (\figone\,\bfA). Through Eq.~\ref{eq:g2}, the real space correlations in these limiting cases appear in photon statistics as bunching $g^{(2)}_\text{ph}(0) > 1$ and antibunching $g^{(2)}_\text{ph}(0) < 1$, respectively (\figone\,\bfB). A full derivation of Eq. 1 is given in the SI.
%Realizing PCM requires a system that is optically active, strongly interacting, and electrostatically tunable. 

We demonstrate PCM in a 1D ensemble of dipolar excitons formed at the atomically sharp lateral interface between \Mo\, and \W\, monolayers \cite{Duan2014,Sahoo2018,Lau2018,Rosati2023,Vandoolaeghe2025}. Unlike conventional 2D monolayer excitons, the type-II lateral band alignment at this junction localizes the conduction-band electron on the \Mo\, side and the valence-band hole on the \W\, side, producing a bound state whose constituent charges are spatially separated across the interface. The resulting interfacial excitons are quantum confined to 1D along the heterojunction and carry an exceptionally large, permanent in-plane electric dipole moment ($|\vec{p}| = e\times 2.2\,$nm), along with long radiative lifetimes ($\tau_0 \sim 10\,$ns) and high 1D mobilities along the interface \cite{Vandoolaeghe2025}. Their extreme sensitivity to in-plane electric fields and charge doping, in turn, enables lithographically defined longitudinal trapping potentials, yielding a solid-state realization of 1D dipolar bosons with mesoscopic lengths. Full details of the system, fabrication, and characterization are given in the SI and in Ref.\,\cite{Vandoolaeghe2025}. In what follows, we first establish gate-defined mesoscopic confinement of dipolar excitons. We then demonstrate a density-driven compressibility crossover from a thermal to an incompressible quasi-crystal phase using power-dependent spectroscopy, and finally show its direct signature in the photon statistics of the emitted light. \\

\begin{figure*}[!t]
\centering
\includegraphics[]{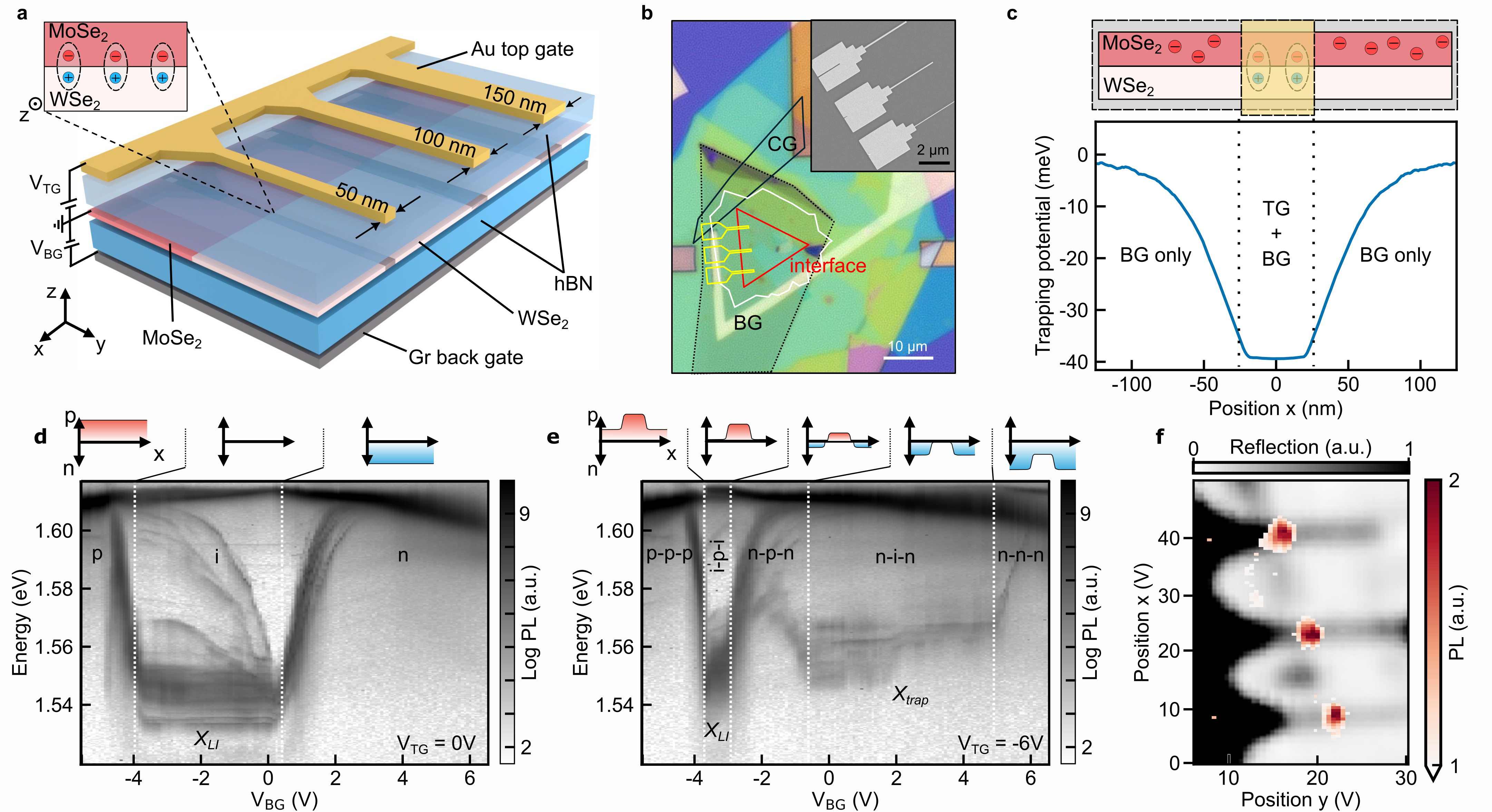}

\caption{\textbf{Mesoscopic confinement of dipolar excitons.} (\bfA) Schematic of the dual-gated device: a lateral \Mo--\W\, semiconductor heterostructure encapsulated in hBN. Nanostructured top-gate fingers of width $L = 50,\, 100,\, 150\,$nm, oriented perpendicular to the lateral interface, define confinement regions for interfacial dipolar excitons (inset). (\bfB) Micrograph of the device with top-gate fingers (yellow) and the lateral interface (red) outlined; inset: SEM image of the top-gate fingers. (\bfC) Simulated longitudinal trapping potential for $L \approx 50\,$nm shows flat-bottomed 1D confinement. (\bfD) $\vbg$-dependent PL spectra at $\vtg = 0$. The unconfined interface exciton $X_\text{LI}$ exhibits a characteristic blueshift upon doping; the doping configuration is indicated above. (\bfE) $\vbg$-dependent PL spectra at $\vtg = -6\,$V. As $\vbg$ dopes the system from $n$ to $p$, the device passes through different spatial doping configurations. In the $n$-$i$-$n$ regime ($-1\,\mathrm{V} < \vbg < 5\,$V), new discrete states emerge in the energy window of $X_\text{LI}$, which is itself blueshifted away by the surrounding doping. (\bfF) Spatially resolved reflection intensity from the device at $755\,$nm, overlaid with a PL scan map recorded at $\vbg = 5\,$V, $\vtg = -6\,$V at the $X_\text{trap}$ energy: emission from the $X_\text{trap}$ states originates exclusively beneath the finger gates, confirming longitudinal confinement.
}
\label{fig:two} 
\end{figure*}

\section*{Entering the mesoscopic regime}
Resolving photon correlations of many-body systems requires reaching the mesoscopic regime, where matter correlations and fluctuations contribute detectable contrast to the emitted light. Two constraints follow from Eq.~\ref{eq:g2}. From the \emph{fluctuation perspective}, a large population $\braket{N}$ dilutes relative fluctuations: even for a perfectly number-stabilized state ($\varn = 0$), the deepest antibunching achievable is $g^{(2)}_\text{ph}(0) = 1 - 1/\braket{N}$ --- the Poisson floor set by the finite particle number. Equivalently, from the \emph{correlation perspective}, the sampling length $L$ sets the window over which $g^{(2)}_\text{mat}(x,x')$ is integrated: when $L$ exceeds the matter coherence length $\xi_\text{coh}$, the $g^{(2)}_\text{ph}$ contrast is suppressed by a factor $\xi_\text{coh}/L$ as multiple incoherent cells are sampled. Together these constraints define the mesoscopic ceiling: to resolve $\sim 5\%$ deviations from Poisson statistics, we need  $\braket{N},\, L/\xi_\text{coh} \lesssim 20$.

\begin{figure*}[!t]
\centering
\includegraphics[]{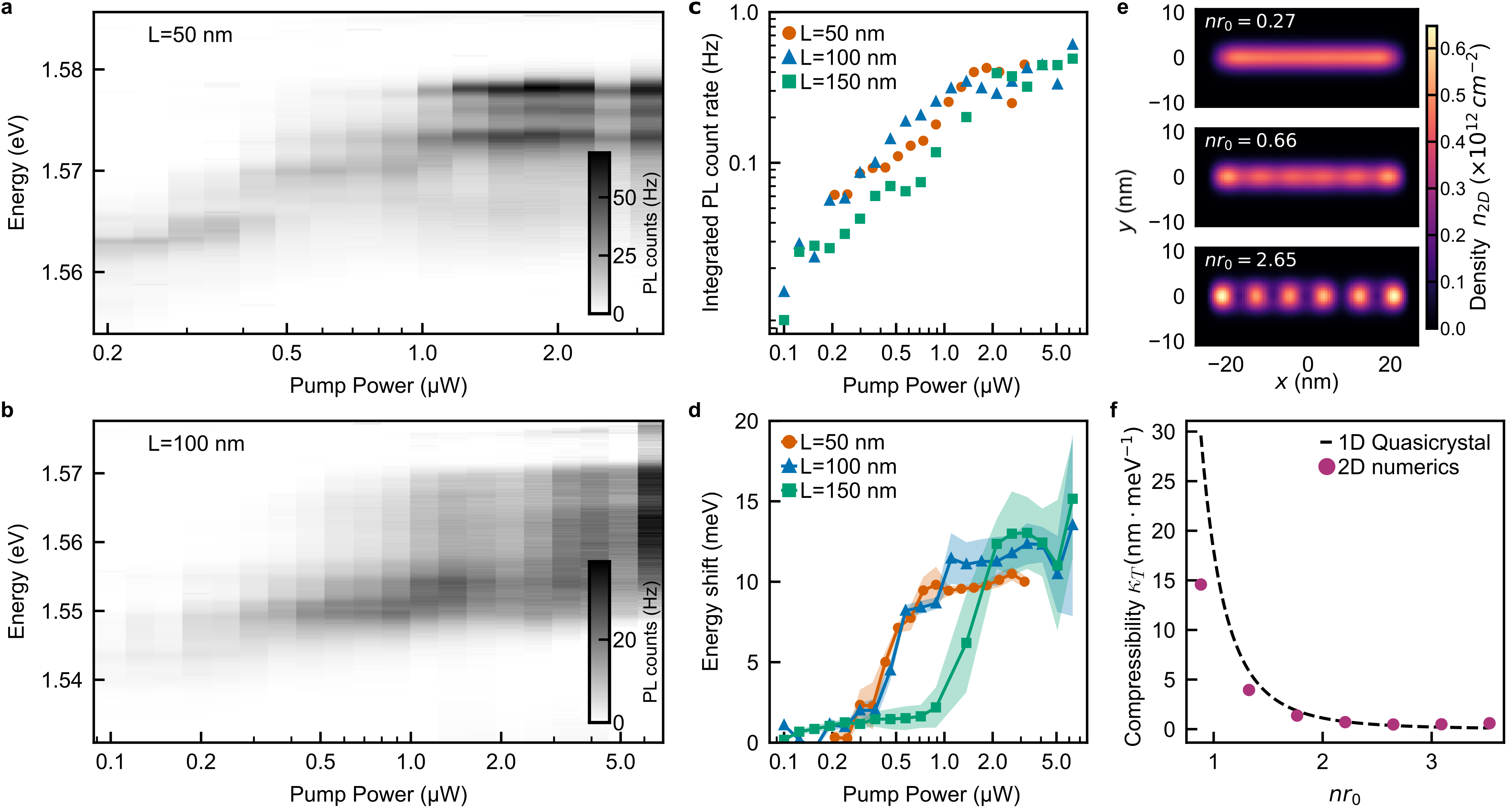}

\caption{\textbf{Interaction-driven crossover.} (\bfA,\,\bfB) Power-dependent PL spectra (normalized by integration time) for the $L = 50\,$nm and $L = 100\,$nm traps, respectively. As the power is increased, emission shifts sequentially from the lowest transverse subband to higher-lying ones. (\bfC) Integrated PL count rate (normalized by the average photon energy) for the $L = 50\,$nm (orange), $L = 100\,$nm (blue) and $L = 150\,$nm (green) traps, displaying a steep rise followed by saturation at higher powers. (\bfD) Mean emission energy for the same traps, exhibiting a similar steep rise and saturation. The integrated counts and the mean energy saturate at the same power $P_\text{sat}$ for each trap, with $P_\text{sat}$ depending on the trap size. Additional data are provided in the SI. (\bfE) Theoretically computed $T=0$ density distribution $n(x)$ in the $50\,$nm trap for increasing values of $nr_0$ ($nr_0 = 0.27$ upper panel, $nr_0 = 0.66$ middle panel, $nr_0 = 2.65$ lower panel), which encompasses the interaction crossover, for the fixed number of particles $\braket{N} = 6$.  (\bfF) Theoretically computed compressibility $\kappa_T = n^{-2}(\partial\mu/\partial n)^{-1}$ as a function of $nr_0$ shows a sharp reduction to zero at $nr_0 \gtrsim 2$ corresponding to $\braket{N} \approx 5$, for $r_0 = 20\,$nm. This is consistent with the analytic expression of 1D dipolar gas compressibility in the quasi-crystal regime (dashed line) \cite{citro2008luttinger}.
}
\label{fig:three} 
\end{figure*}

We reach this mesoscopic regime through gate-defined longitudinal traps for dipolar excitons. Our device consists of a MoSe$_2$--WSe$_2$ lateral heterostructure encapsulated in hexagonal boron nitride (hBN) and stacked on a graphite back-gate (\figtwo\,\bfA). We implement longitudinal confinement through lithographically patterned top-gate fingers ($50, 100,$ and $150\,$nm wide) extending perpendicular to the interface (\figtwo\bfB; SEM image in inset). The trapping mechanism exploits the extreme sensitivity of interfacial excitons to the local electrostatic environment: a back-gate voltage globally dopes the 1D interface, while the top-gate fingers locally counteract this doping, producing depleted, neutral pockets where the exciton energy is lowered. Finite-element electrostatic simulations (\figtwo\,\bfC) confirm flat-bottomed potentials of width set by the top-gate geometry; we treat them as a qualitative guide, given that fringing and device non-idealities can modify the realized trap depths and lengths.

The experimental signature of exciton confinement is evident from the photoluminescence (PL) spectra, acquired from the $50\,$nm channel. Corresponding results from the $100\,$nm and $150\,$nm channels are shown in SI. At $V_{\text{TG}} = 0\,$V (\figtwo\,\bfD), the 1D interface is uniformly doped by the back-gate; we observe the characteristic doping-dependent blueshift of the lateral interface excitons ($X_\text{LI}$) as $\vbg$ is varied  \cite{Vandoolaeghe2025}. At $\vtg = - 6\,$V (\figtwo\,\bfE), the electrostatic landscape is dramatically altered. As $X_\text{LI}$ state blueshifts with global doping, a series of discrete states emerge at the energy of the original neutral interface exciton. 
The key signature of trapping is therefore the appearance of discrete states locked at the $X_\text{LI}$ energy, whereas the surrounding $X_\text{LI}$ state has blueshifted. We label these longitudinally confined states $X_\text{trap}$ hereafter. In \figtwo\,\bfF, we show the reflected intensity at $755\,$nm overlaid with the PL emission spectrally filterd at the $X_\text{trap}$ energy. While the reflection scan directly reveals the underlying finger gate structure of the device, the $X_\text{trap}$ PL is spatially restricted strictly to the finger gate regions. This directly confirms that excitons are confined within the gate-depleted regions, where the local doping has been neutralized. Their spectra retain the multi-subband structure of $X_\text{LI}$, which arises from transverse excited modes (along $y$) of the interface (previously characterized in Ref.\cite{Vandoolaeghe2025}). 

These measurements demonstrate 1D mesoscopic traps with lengths $L \approx 50$, $100$, and $150\,$nm and transverse widths $\ell_y \approx 2-3\,$nm set by the interface width. The calculated longitudinal mode spacing in the smallest trap, $\hbar\omega_x = \hbar^{2}\pi^2/(2m_X L^{2}) \approx 100\,\mu$eV, lies well below the exciton linewidth $\Gamma_X \sim 1\text{--}2\,$meV, yielding continuous motion along $x$; the transverse motion is quantized into resolvable subbands with $\hbar\omega_y \approx 3\text{--}5\,$meV. This combination --- a longitudinal continuum within a discrete transverse spectrum---provides the finite-length quasi-1D geometry required to explore the many-body regime of dipolar excitons.

\section*{Interaction-driven crossover}
We now investigate the interaction-induced energy shifts and compressibility crossover in the trapped 1D dipolar gas using PL spectroscopy. In our driven--dissipative platform, the chemical potential ($\mu$) is controlled by the optical pump power: photogenerated carriers form 2D excitons in the surrounding monolayers, diffuse to the 1D interface, form charge-transfer states and are captured by the gate-defined potential wells. The steady-state occupancy is set by the kinetic balance between this effective drive and radiative decay. Since the thermalization time $\tau_\text{th} \sim 1-10\,$ps and interaction time $\tau_\text{int} \sim \hbar/\mu \sim 1\,$ps (for $\mu \approx 1\,$meV) are both three orders of magnitude faster than the decay timescale ($\Gamma^{-1} \sim 10\,$ns), the system reaches a quasi-equilibrium steady state between emission events.

The 1D dipolar excitonic system is a particularly powerful testbed for the PCM framework. The many-body physics of 1D dipolar bosons is captured by the Luttinger liquid theory \cite{Cazalilla2011,Deuretzbacher2010,Roscilde2010,Peng2024}. The dipolar interaction between excitons $V(x) \sim C_{dd}/x^{3}$ (where $C_{dd} = e^{2}d^{2}/4\pi\epsilon_{0}\epsilon$) defines a characteristic dipolar length $r_0 = m_X C_{dd}/\hbar^{2} \sim 20\,$nm, and the interaction to kinetic energy ratio $n r_0 = E_\text{int}/E_\text{K}$ which grows linearly with density, where
$E_\text{int} \sim C_{dd} n^{3}$ and $E_\text{K} \sim \hbar^{2} n^{2}/m_X$ ($m_X$ is exciton mass). Physically, the cubic dipolar interaction scaling outpaces the quadratic kinetic scaling, so increasing density progressively tips the balance from kinetic to interaction energy. In the dilute regime ($n r_0 \ll 1$), excitons rarely encounter each another and the gas behaves as a weakly interacting thermal ensemble; at higher densities ($n r_0 \gtrsim 1$), the gas is expected to cross smoothly through a Luttinger superfluid phase into the strongly correlated quasi-crystalline state \cite{Citro2007, Sinha2007, Roscilde2010, Deuretzbacher2010, Peng2024}. Tuning the exciton density thus drives the system continuously from a compressible thermal phase to a number-stabilized correlated state within a single device. 

\figthree\,\bfA\, and \bfB\, show the continuous wave (CW) excitation power-dependent PL spectra for the $50\,$nm and $100\,$nm trap. We observe a striking sequential loading of the transverse subbands of the dipolar wire in both systems. At low power, emission originates primarily from the lower subbands. As the power is increased, each subband in turn blueshifts steeply --- dipolar repulsion raises the cost of each additional exciton --- and then saturates in energy when $\mu$ climbs past $\hbar\omega_y$, the transverse mode spacing. Beyond this point, additional excitons are forced into the next transverse mode. At the highest powers, the uppermost populated subband carries the largest fraction of the signal. The sequential filling visible in \figthree\,\bfA\, is therefore a spectroscopic readout of the stiffening equation of state of the confined dipolar excitons ($X_\text{trap}$).

The fits of these power-dependent spectra simultaneously deliver two thermodynamic observables. The integrated PL counts (\figthree\,\bfC, log-log scale), for the $50\,$nm (orange), $100\,$nm (blue) and $150\,$nm (green) traps, trace the steady-state particle number $\braket{N}$, while the fitted mean emission energy shift (\figthree\,\bfD, lin-log scale) tracks the effective chemical potential $\mu$. Both exhibit a steep rise at low power followed by a sharp saturation at $P_\text{sat} \sim 0.7\text{--}2\,\mu W$, with $\mu$ blueshifting by $\sim 10-15\,$meV across this range. The simultaneous saturation of $\braket{N}$ and $\mu$ in each trap identifies a sharp drop in compressibility $\kappa_T = n^{-2}(\partial \mu/\partial n)^{-1}_T$ across $P_\text{sat}$: beyond this power, the matter phase resists further densification --- in marked contrast to typical 2D excitonic systems, where the total PL grows approximately linearly with pump power up to extremely high densities. 

To confirm that this observed saturation reflects genuine reduction of compressibility, we compute the zero temperature equilibrium properties of the trapped 1D dipolar excitons using the multi-configuration time-dependent Hartree method (SI) (\figthree\,\bfE\, and \bfF). \figthree\,\bfE\, shows the density distribution $n(x)$ for different values of $nr_0$, which encompasses the crossover, for a fixed particle number $\braket{N} = 6$ (by varying $r_0$). As $nr_0$ increases from $\approx 0.3$ to $\approx 2.6$, the gas evolves from a delocalized, kinetic-energy-dominated regime into a quasi-crystalline, interaction-dominated state. \figthree\,\bfF\, shows the corresponding numerically calculated isothermal compressibility $\kappa_T = n^{-2}(\partial \mu/\partial n)^{-1}$, which drops sharply toward zero beyond $nr_0 \gtrsim 1$, in agreement with the analytic 1D dipolar crystal expectation~\cite{citro2008luttinger}. These $T = 0$ simulations address the high-density side of the crossover; the low-density thermal regime lies outside their scope. Nevertheless, they confirm the central picture: as density grows, the 1D dipolar exciton gas approaches an incompressible state with quasi-crystalline spatial order. Further details are given in the SI.

This nonlinear equation of state provides the thermodynamic foundation for the suppression of photon fluctuations. For this quasi-equilibrium system, through Eq.\,\ref{eq:g2}, this collapse of $\kappa_T$ must imprint directly on the photon statistics as a drop of $g^{(2)}_\text{ph}(0)$ below unity — a prediction we test next.

\begin{figure*}[ht!]
\centering
\hspace*{-0.3cm}
\includegraphics{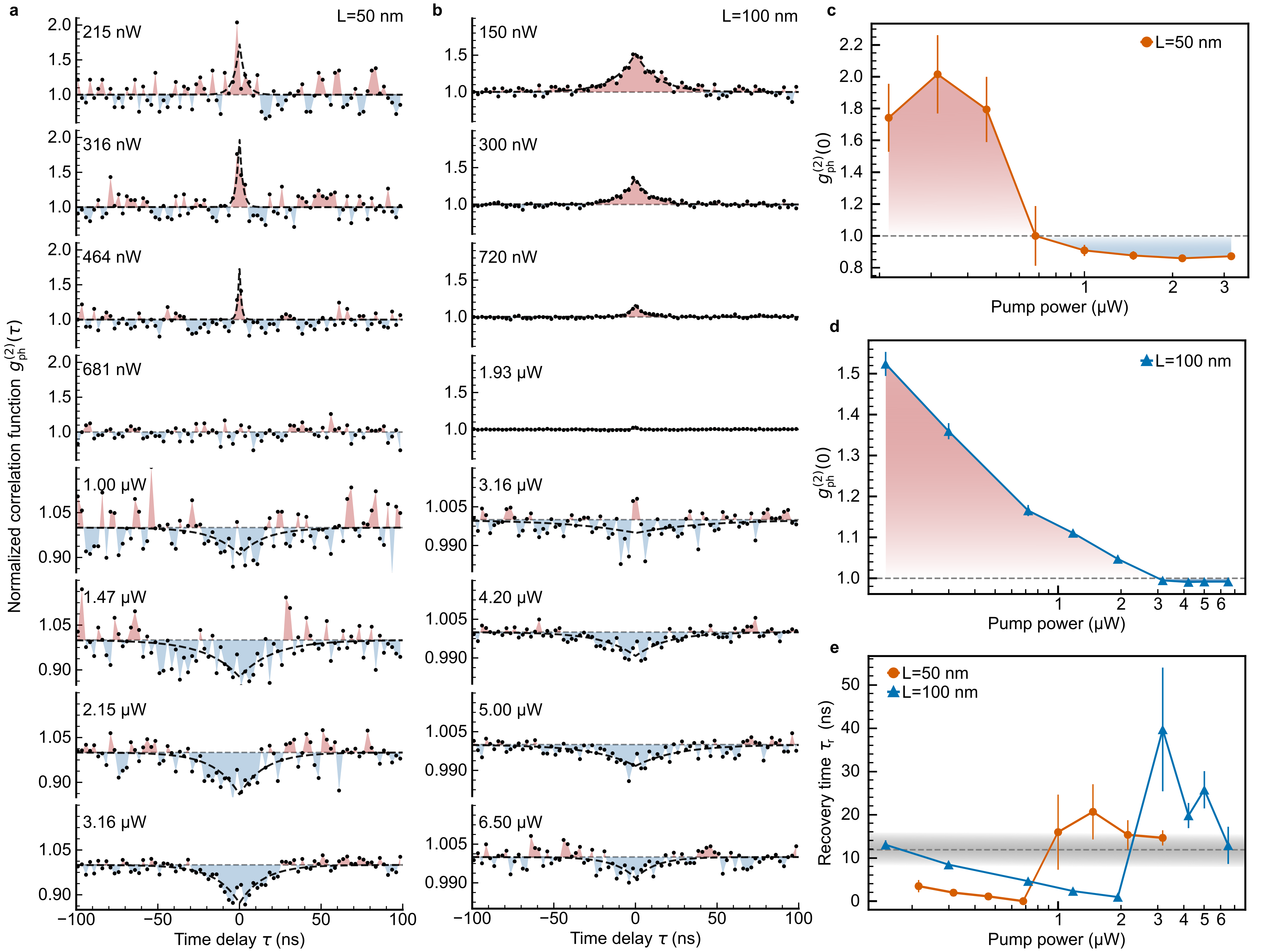}

\caption{\textbf{Many-body blockade of photon emission.}
(\bfA,\,\bfB) Power-dependent $g^{(2)}_\text{ph}(\tau)$ for the $L = 50\,$nm and $100\,$nm traps. In both, $g^{(2)}_\text{ph}(\tau)$ evolves continuously from bunching at low power to antibunching at high power --- a many-body crossover qualitatively distinct from the always-antibunched response of single quantum emitters such as quantum dots and defects.
(\bfC,\,\bfD) Extracted $g^{(2)}_\text{ph}(0)$ vs power. The $50\,$nm trap traverses from the single-mode chaotic limit $g^{(2)}_\text{ph}(0) \approx 2$ to a strongly antibunched value $g^{(2)}_\text{ph}(0) \approx 0.84$. In the $100\,$nm trap, both the bunching contrast and the antibunching dip are suppressed ($\approx 1.5 \to 0.99$), in agreement with the mesoscopic-confinement scaling of Eq.~\ref{eq:g2}.
(\bfE) Recovery time $\tau_r$ vs.\ power for both traps. Starting at $\tau_r \approx 5\text{--}10\,$ns at low power, $\tau_r$ decreases as the system approaches saturation and then rises sharply with the onset of antibunching.
Together, these measurements unambiguously demonstrate the many-body photon blockade --- a continuous crossover from a weakly interacting 1D Bose gas to a strongly correlated, number-stabilized excitonic fluid --- and establish photon correlation microscopy as a quantitative probe of strongly correlated quantum matter.}
\label{fig:four} 
\end{figure*}

\section*{Many-body blockade of photon emission}
To investigate the emergence of many-body correlations, we perform power-dependent photon correlation spectroscopy using a Hanbury Brown--Twiss (HBT) interferometer. The PL is spectrally filtered to collect photons emitted from all transverse subbands identified in \figthree\,\bfA\, and \bfB, so that the measurement captures the collective statistics of the ensemble rather than those of any single subband.

Figs.\,\ref{fig:four}\,\bfA\, and \bfB\, show the measured second-order photon correlation function $g^{(2)}_\text{ph}(\tau)$ for the $50\,$nm and $100\,$nm traps, respectively, across CW excitation powers spanning the thermodynamic crossover of \figthree. As power is increased, both traps undergo a continuous transition --- from a strong bunching peak at low power, through the Poissonian regime $g^{(2)}_\text{ph}(0) = 1$, to a clear antibunching dip at high power. Similar qualitative behavior is observed in the $150\,$nm trap (SI), with diluted contrast as expected from Eq.\,\ref{eq:g2}. The observation of antibunching at high powers is a model-independent signature of sub-Poissonian matter statistics in the 1D dipolar excitonic ensemble, and constitutes the central manifestation of the many-body blockade.  

The zero-delay value $g^{(2)}_\text{ph}(0)$ extracted from these data is summarized in \figfour\,\bfC\, ($50\,$nm) and \bfD\,($100\,$nm). At low power, the $50\,$nm trap saturates at $g^{(2)}_\text{ph}(0) \approx 2$ --- the chaotic-thermal limit of a thermal Bose gas occupying a single coherence cell. The $100\,$nm trap reaches $g^{(2)}_\text{ph}(0) \approx 1.5$ at low power, still strongly bunched but reduced as the trap begins to sample more than one coherence cell; through Eq.~(\ref{eq:g2}), the two values together pin the matter coherence length at $\xi_\text{coh} \sim 50\,$nm. 

As power is increased, both traps cross to antibunching at the same $P_\text{sat}$ identified spectroscopically in \figthree --- the bunching-to-antibunching transition occurs synchronously with the collapse of compressibility. At the highest powers, $g^{(2)}_\text{ph}(0) = 0.84 \pm 0.03$ in the $50\,$nm trap and $g^{(2)}_\text{ph}(0) = 0.99 \pm 0.001$ in the $100\,$nm trap. The $L = 50\,$nm value essentially saturates the Poisson floor $1 - 1/\braket{N}$ for $\braket{N} \approx 5\text{--}6$, indicating a fully number-stabilized state consistent with \figthree. The $100\,$nm trap antibunches more weakly: with $\braket{N} \approx 10-20$, the observed value lies above the Poisson floor $g^{(2)}_\text{ph}(0) \approx 0.9-0.95$, placing it in a partially number stabilized regime.

Across this crossover, the measurements trace three regimes of the 1D dipolar excitonic gas, following Eq.\,\ref{eq:g2}: a thermal regime at low power ($nr_0 \ll 1, n\lambda_T \lesssim 1$) where $g^{(2)}_\text{ph}(0)$ probes the coherence length $\xi_\text{coh}$; a quantum degenerate regime  ($n\lambda_T \gtrsim 1$) where $g^{(2)}_\text{ph}(0)$ passes unity; and a strongly correlated quasi-crystal regime ($nr_0 \gtrsim 1, n\lambda_T > 1$) where dipolar repulsion drives sub-Poissonian statistics \cite{Peng2024}. 

Beyond the magnitude of $g^{(2)}_\text{ph}(0)$, the temporal structure of $g^{(2)}_\text{ph}(\tau)$ provides an additional observable -- the recovery time $\tau_r$ over which Poissonian statistics is restored. While $g^{(2)}_\text{ph}(0)$ exhibits a monotonic dependence on power, $\tau_r$ shows a non-monotonic evolution in both traps (\figfour\,\bfE). Specifically, $\tau_r$ decreases with power, reaches a minimum at the bunching-to-antibunching crossover, and rises sharply at high power, eventually exceeding the radiative lifetime $\tau_0 \sim 10\,$ns (SI). At low power, $\tau_r$ approaches $\tau_0$: fluctuations relax through single-particle decay. The rise well beyond $\tau_0$ in the high density regime is qualitatively different: density fluctuations relax on a timescale set by interactions rather than by single-particle decay. This dynamical signature, observed in lockstep with the spectral and correlation-magnitude crossovers, provides an additional, independent line of evidence that the bunching-to-antibunching evolution reflects a thermodynamic crossover of the underlying matter, inaccessible to any independent-emitter picture. 

%The bunching-to-antibunching crossover observed in \figfour\ traces the excitonic system through three distinct regimes of the 1D dipolar Bose gas, each with a direct photon-statistical signature. At the lowest powers ($n r_0 \ll 1$, $n\lambda_T \lesssim 1$), the system is a weakly interacting thermal Bose gas. The bunching contrast itself measures the matter coherence length: the saturation at $g^{(2)}_\text{ph}(0) \approx 2$ in the $50\,$nm trap, together with the reduced value $\approx 1.5$ in the $100\,$nm trap, pins $\xi_\text{coh} \sim 50\,$nm (\figfour\,\bfC,\bfD). This corresponds, via the thermal de Broglie relation, to an effective temperature $T_\text{eff} \approx 1.7\,$K (SI) --- well above the cryostat temperature, as expected for an optically driven gas. As the power is increased, the gas crosses into the quantum-degenerate regime ($n\lambda_T > 1$), where number fluctuations approach the Poissonian limit $\varn = \braket{N}$ and $g^{(2)}_\text{ph}(0)$ passes through unity. Further densification drives the system into the strongly interacting regime ($nr_0 > 1$), where dipolar repulsion stiffens the gas, $g^{(2)}_\text{ph}(0)$ drops below unity, and the matter pair correlator develops a quasi-crystalline correlation hole. 

%%___

\section*{Discussion}
We have demonstrated \emph{photon correlation microscopy} (PCM) using a mesoscopically confined ensemble of 1D dipolar excitons as a testbed. While reduction in number fluctuations have been inferred from PL intensity statistics in related platforms -- notably in interlayer excitons in electrostatic lattices \cite{Lagoin2024, Lagoin2022Mott} -- we show that 
non-classical light can directly emerge from correlated matter. As the density is tuned in our system, the emitted light evolves continuously from chaotic-light bunching ($g^{(2)}_\text{ph}(0) \approx 2$) to non-classical antibunching ($g^{(2)}_\text{ph}(0) \approx 0.84$), a model-independent optical readout of the matter crossing from a compressible thermal Bose gas into a strongly correlated, number-stabilized fluid. The accompanying non-monotonic evolution of the recovery time $\tau_r$ independently signals the onset of collective, interaction-set relaxation. Together these establish a \emph{many-body photon blockade}: non-classical photon statistics emerging collectively from interactions in quantum matter.

This departs from the isolated-emitter paradigm of quantum optics. While photon antibunching in atoms, quantum dots, and color centers is a fixed constraint of an individual emitter's level structure, the blockade here is tunable in situ through density, geometry, and interaction strength. Such control is precisely what electrostatically gate-defined excitonic platforms have sought --- both in GaAs coupled quantum wells and in TMD heterostructures \cite{Schinner2013, Thureja2022, Hu2024, Heithoff2024, Thureja2024} --- yet non-classical light from such systems had not been realized. It is achieved here because the gate-defined trap combined with 1D dipolar excitonic interactions result in a strongly correlated state within the mesoscopic regime. This points toward a new kind of quantum-light source: one whose statistics are stabilized by interactions rather than by isolation, and therefore in principle less susceptible to spectral diffusion, blinking, and dephasing that limit individual emitters.

For many-body physics, our work introduces a new class of optical probe that provides direct, non-invasive access to the four-point density-density correlator of matter. It is restricted neither to dipolar excitons nor to bosons (SI). The roles of matter and emitter can in principle be decoupled: while the dipolar excitons here double as the photon source, the same mapping applies whenever an optical excitation co-exists with, or is dressed by, a strongly correlated electronic system and reports its local density. Trions -- excitons bound to a single charge carrier -- are natural candidates as their emission rate tracks the local charge density of the underlying electron system, providing a direct readout of electronic correlations. This potentially extends PCM to correlated phases in vdW heterostructures \cite{Tang2020, Smolenski2021}, where the local four-point correlator carries signatures of spatial, spin and topological order that are challenging to access via bulk-averaged transport and spectroscopic probes. Promising targets include electronic and excitonic Mott insulators and generalized Wigner crystals in moir\'e superlattices \cite{Regan2020} and fractional quantum Hall and Chern insulators \cite{Li2026}. Moreover, the required mesoscopic sampling need not come from confinement alone: near-field probes \cite{Beams2014}, plasmonic nanoantennas, or sub-wavelength photonic modes could restrict the sampled region within macroscopic ensembles. More broadly, PCM brings to solid-state correlated matter a class of microscopic correlation measurements previously accessible only in ultracold atom systems via quantum gas microscopy \cite{Bakr2009, Hilker2017}. 

Several directions follow. A complete microscopic theory of the many-body blockade across these systems --- including the strongly correlated 1D dipolar exciton fluid itself --- remains open. The long radiative lifetime of dipolar excitons positions PCM to track the real-time formation of the correlation hole after a quench, a window onto thermalization in a driven-dissipative correlated fluid. More broadly, polarization-, frequency-, and spatially-resolved correlation measurements could extend access to spin, valley, and non-local correlations beyond the pair level, in line with recent theory proposals for cavity-embedded systems \cite{Kass2024} and photon scattering in quantum materials \cite{Nambiar2025}. This bridges quantum optics and condensed matter: non-classical light becomes a probe of strongly correlated matter, and strongly correlated matter becomes a resource for generating non-classical light.

\textbf{Acknowledgements.} We thank Yoshihisa Yamamoto, Tony F. Heinz, Tilman Esslinger, Tobias Donner, Ajit Srivastava, Paolo Molignini and Lukas Novotny for insightful discussions.

\noindent
\textbf{Funding.} This work was supported by Swiss National Science Foundation (SNSF) Starting Grant no. 211448. P.S. acknowledges the Department of Science and Technology (DST) (Project Code: DST/NM/TUE/QM-1/2019 and National Quantum Mission (NQM) DST/QTC/NQM/QMD/2024/4/(G)), India. K.W. and T.T. acknowledge support from the JSPS KAKENHI (grant numbers 19H05790, 20H00354, and 21H05233).
N.D. acknowledges funding from the Swiss National Science Foundation (SNSF) grant numbers 200021--207537 and 200021--236722, by the Deutsche Forschungsgemeinschaft (DFG, German Research Foundation) under Germany's Excellence Strategy EXC2181/1-390900948 (the Heidelberg STRUCTURES Excellence Cluster) and and the Swiss State Secretariat for Education, Research and Innovation (SERI). \\

\textbf{Author contributions.}
P.A.M. and T.C. originated the PCM concept and established the experimental framework; E.V. and I.L. developed the device architecture, fabricated the lateral heterostructure devices, performed the primary spectroscopic measurements and data analysis, with inputs from P.A.M; C.V. and T.C. performed the photon correlation measurements and spectroscopy, and analyzed the $g^{(2)}$ statistics; K.W. and T.T. provided the hBN crystals. S.S. performed the theoretical analysis of the 1D dipolar exciton gas under the supervision of N.D.; P.R. designed, grew and characterized the lateral heterostructure materials under the supervision of P.K.S., who led the material platform development enabling this study; T.C. and P.A.M. jointly led and supervised the device engineering and optics experiments; All authors contributed to the preparation of the manuscript.

\bibliography{Reference}

% ═══════════════════════════════════════════════════════════════════════════
%  SUPPLEMENTARY INFORMATION
%  Appears as an independent document glued after the main text.
% ═══════════════════════════════════════════════════════════════════════════
\clearpage
\newpage

% ── Switch to S-prefixed numbering ────────────────────────────────────────
\renewcommand{\theequation}{S\arabic{equation}}
\renewcommand{\thefigure}{S\arabic{figure}}
\renewcommand{\thetable}{S\arabic{table}}
\renewcommand{\thesection}{S\arabic{section}}

% ── Reset all counters so numbering starts from S1 ────────────────────────
\setcounter{equation}{0}
\setcounter{figure}{0}
\setcounter{table}{0}
\setcounter{section}{0}
\setcounter{page}{1}

% ── Supplementary title block ─────────────────────────────────────────────
\begin{center}
  {\Large\bfseries Supplementary Information\\[4pt]
  ``Photon correlation microscopy of quantum matter''}\\[10pt]
  Elie Vandoolaeghe$^{*,1}$,
  I\~{n}igo Lasheras$^{*,1}$,
  Chirag Vaswani$^{2}$,
  Sampriti Saha$^{3}$,
  Purbasha Ray$^{4}$,
  Takashi Taniguchi$^{5}$,
  Kenji Watanabe$^{5}$,
  Prasana Sahoo$^{4}$,
  Nicolo Defenu$^{3}$,
  Thibault Chervy$^{\dagger,2}$,
  Puneet A. Murthy$^{\ddagger,1}$\\[6pt]
  {\small
    $^{1}$Institute for Quantum Electronics, ETH Z\"urich, CH-8093 Z\"urich, Switzerland\\
    $^{2}$NTT Research, Inc.\ Physics \& Informatics Laboratories, 940 Stewart Dr, Sunnyvale, CA 94085\\
    $^{3}$Institute for Theoretical Physics, ETH Z\"urich, CH-8093 Z\"urich, Switzerland\\
    $^{4}$Quantum Materials and Device Research Lab, Materials Research Center,
           Indian Institute of Technology, Kharagpur, India\\
    $^{5}$National Institute for Materials Science, Namiki 1-1, Tsukuba, 305-0044, Ibaraki, Japan\\[4pt]
    $^{*}$These authors contributed equally\quad
    $^{\dagger}$\texttt{thibault.chervy@ntt-research.com}\quad
    $^{\ddagger}$\texttt{murthyp@ethz.ch}
  }
\end{center}
\vspace{6pt}
\hrule
\vspace{12pt}

\tableofcontents
\section{Experimental details}
\subsection{Experimental setup}
Optical spectroscopy measurements were carried out in a dilution refrigerator (Oxford Proteox) at a nominal base temperature of 40mK. A tunable continuous-wave Ti:sapphire laser (Spectra Physics Matisse) served as the excitation source at 730 nm and a cryogenic high numerical aperture(NA) objective (NA = 0.8, Attocube systems) was used for excitation and collection. Excitation power was controlled using a variable optical attenuator (VOA). For spectrally resolved measurements, the collected emission was filtered and guided to a spectrometer (Andor Shamrock 750) with a cooled CCD camera (Andor Newton 940).   

\subsection{Correlation measurements}
Photon correlation measurements were carried out in a fiber-based Hanbury-Brown Twiss (HBT) setup coupled with superconducting nanowire single photon detectors (SNSPDs, Single Quantum). The collected emission was spectrally filtered around the wavelength of interest before coupling to the HBT setup. Detection events on the two SNSPDs were recorded by a time-tagger system (Swabian instruments) with a temporal jitter of 5ps. 

For fluorescence lifetime measurements (time-correlated single photon counting, TCSPC), a pulsed light source (NKT Photonics SuperK Evo, 1ps pulse duration, 730nm central wavelength, 20MHz repetition rate) was used for excitation. Part of the pump light was sent to one of the SNSPD detectors to provide a 'start' signal, and the X$_{\mathrm{trap}}$ PL was sent to the second detector to provide the 'stop' signal. Start-stop histograms were recorded at different pump powers, and fitted using a single exponential tail-fit model. Representative data are shown in the Additional data section below.

\subsection{Device fabrication and details}
The device studied in the main text consists of a charge tuneable lateral heterostructure (LHS) of MoSe$_2$/WSe$_2$ monolayer electrically contacted with a few-layer graphene flake, encapsulated between two hBN flakes, on top of another few-layer graphene flake which act as the bottom gate. 

This vertical heterostructure was assembled using a standard dry-transfer stacking technique: first, a polydimethylsiloxane (PDMS) stamp coated with a polycarbonate (PC) film was used to pick up each flake sequentially; next, the stack was contacted by depositing it on top of pre-patterned gold electrodes on a Si/SiO$_2$ substrate; finally the residual PC was dissolved in chloroform. After assembly, the 13 nm-thick finger gates with lateral widths of 50, 100, and 150 nm were patterned on top with electron-beam lithography, followed by gold deposition and bilayer lift-off. For further information about the photoluminescence (PL) spectroscopy of such LHS and its properties, see \cite{Vandoolaeghe2025}.

\section{Additional data}
\subsection{Additional data for the confinement of excitons}

To characterize the electrostatic response of the sample, identify the charge neutrality condition and find the right voltage configuration for the trapping of lateral interface (LI) dipolar excitons, we perform gate-dependent PL measurements at the intersection of the finger gates with the MoSe$_2$/WSe$_2$ interface. Given the diffraction limited size of our optical spot ($\approx 0.7\,\mu$m), both the region under the finger gate and the region outside of it are illuminated simultaneously. We can label the doping configurations in each region (n, p or i) from left to right along the interface. Trapping occurs when the charge neutrality condition is met only in the region beneath the finger, the n-i-n case, since the LI excitons strongly blueshift as their environment gets doped. For the i-n-i case, neutral LI excitons are pumped outside the finger and are thus unconfined.
\begin{figure}[!htbp]
\centering
\includegraphics[width=\linewidth]{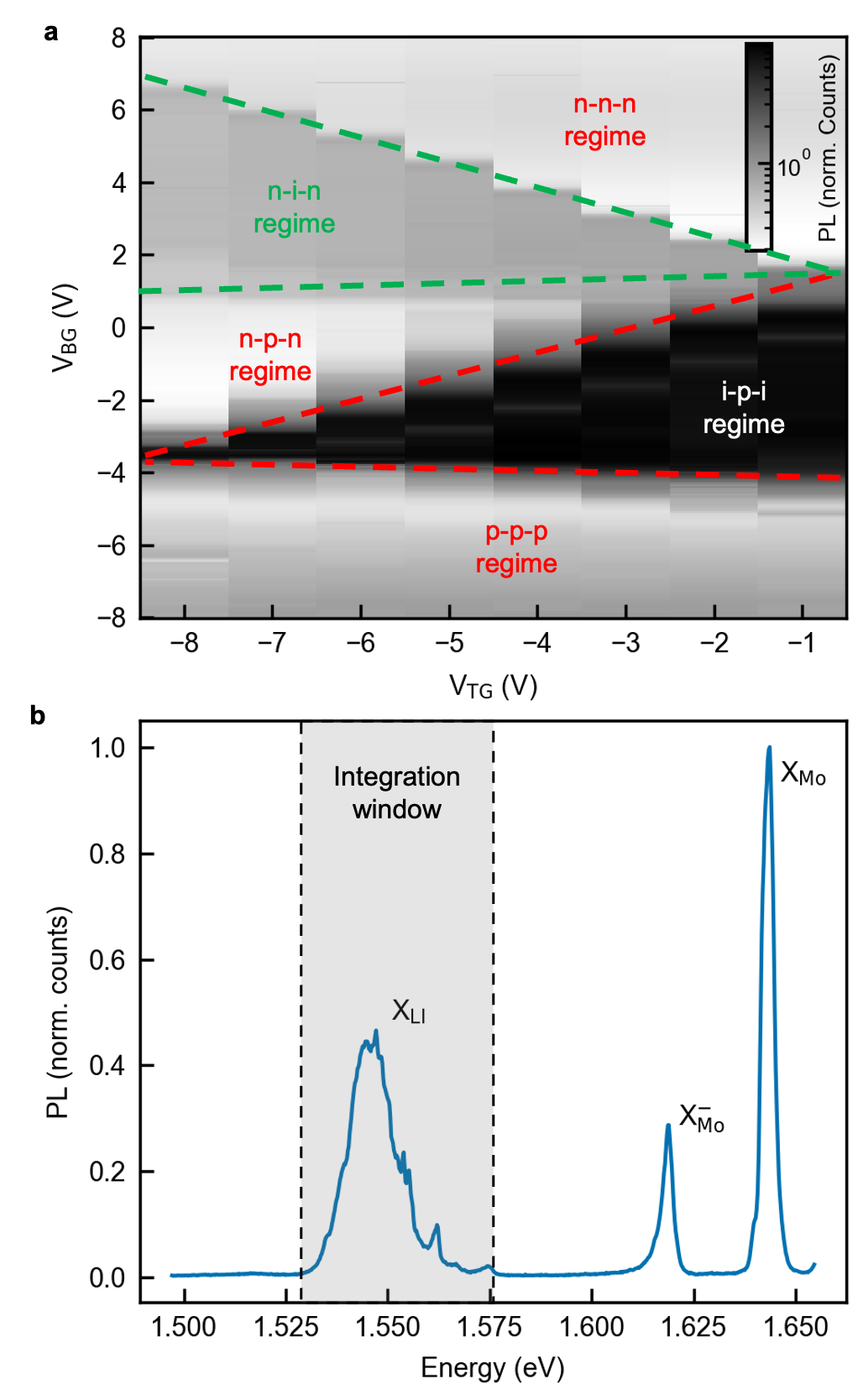}
\caption{\textbf{Full gate-dependent integrated PL map for the 100nm finger}. \textbf{(a)} PL intensity map as a function of both $\vbg$ and $\vtg$ voltages, the color map representing PL spectra integrated over the specific energy range corresponding to X$_{\mathrm{LI}}$ in the neutral regime (see panel \bfB). Sections of high intensity correspond to the appearance of a charge neutrality region in the sample. We highlight the transitions from the different doping regimes with dashed lines, with the n-i-n confinement configuration in green. \textbf{(b)} PL spectrum emitted by the 100 nm finger region at $\vtg =0 V$ and $\vbg = 0V$. The gray area shows the energy integration window to obtain the colormap shown in \bfA.}
\label{fig-sup:PLmap}
\end{figure}

\begin{figure*}[!htbp]
\centering
\includegraphics{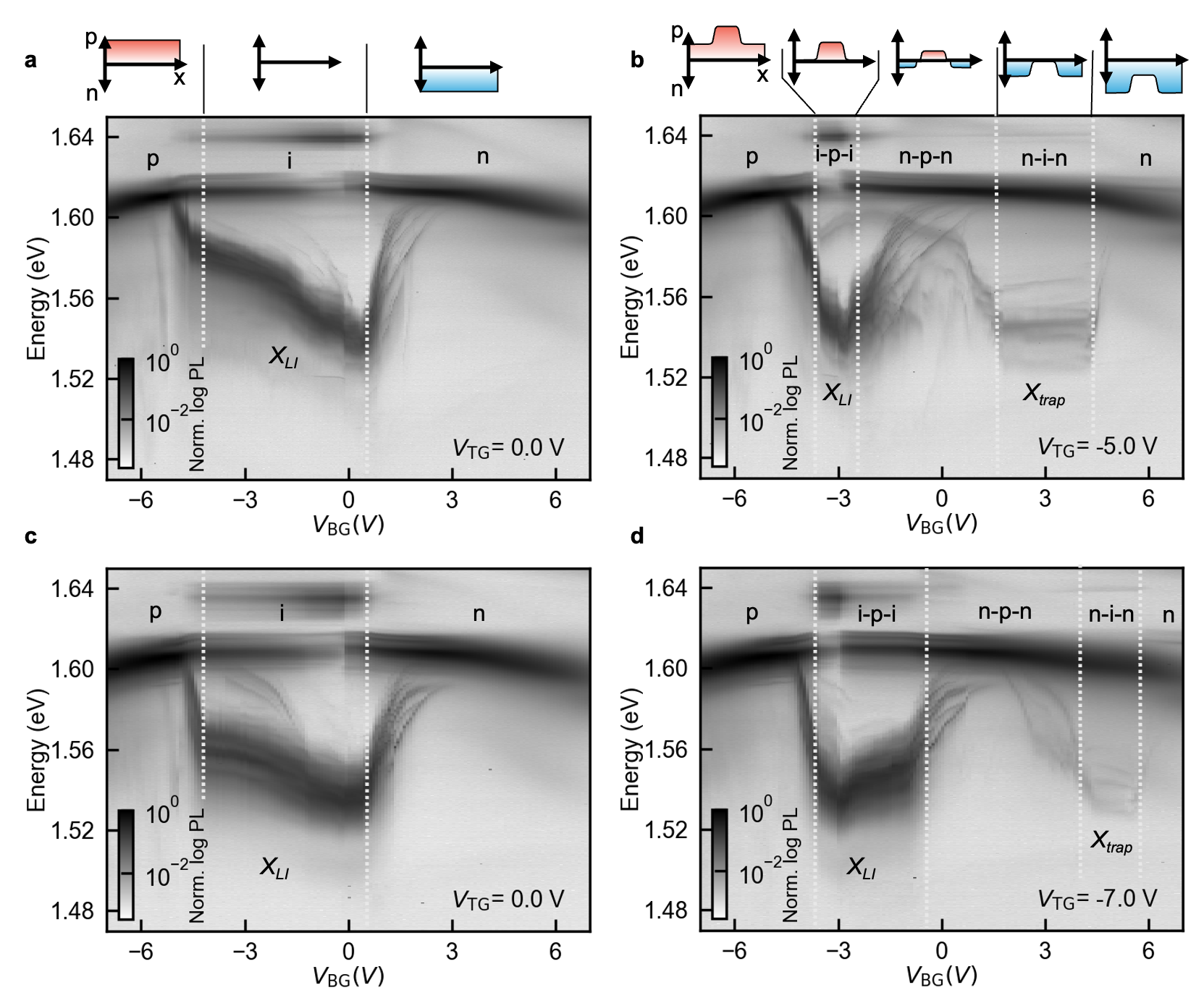}
\caption{\textbf{Gate-dependent PL Spectra for the other fingers}. \textbf{(a)} PL spectra for the 100\,nm finger as a function of $\vbg$ with $\vtg = 0V$, showing the blue shift of the LI exciton with doping, which leads to a steep shift ($\approx 70\,$ meV within 1\,V voltage range). \textbf{(b)} PL spectra for the 100\,nm finger as a function of $\vbg$ with $\vtg = -5V$, where a new feature appears around $\vbg = 3V$, corresponding to the confined excitons, X$_{\mathrm{trap}}$. These X$_{\mathrm{trap}}$ have a similar doping dependence than the unconfined ones, but appear neutral in a voltage window which has blue shifted strongly the rest on the interface states. \textbf{(c)} and \textbf{(d)} present the same data for the 150 nm wide finger, with a similar interpretation.}
\label{fig-sup:PL_otherfingers}
\end{figure*}

Fig.~\ref{fig-sup:PLmap}\bfA\, shows a two-dimensional map of the spectrally integrated PL intensity of the neutral-environment LI exciton (X$_{\mathrm{LI}}$) as a function of both the back-gate voltage $\vbg$ and the top-gate voltage $\vtg$, for the 100 nm width finger gate.  Sections of high PL intensity in the map correspond to the appearance of a charge-neutrality region under the optical spot, where the interface exciton emission is bright. This allows us to clearly delineate the boundaries between the different doping configurations, indicated by the dashed lines. Within this map, we further identify the sections resulting in the n-i-n confinement, highlighted in green, which host the relevant configurations for the discussion in the main text. The number of counts collected from this region is an order of magnitude smaller than the one coming from the i-p-i regime, consistent with the aspect ratio between the finger width and the diffraction limited optical spot. Similar color maps can be constructed for the 50 and 150 nm width fingers, showing the same doping configurations. An example PL spectrum acquired at $\vtg$=0 V and $\vbg$=0 V is shown in Fig.~\ref{fig-sup:PLmap}\bfB\,, with the gray shaded area indicating the spectral integration window used to construct the intensity map in panel \textbf{a}. 

Following the same structure of panels \textbf{d} and \textbf{e} in Fig.2 of the main text, we plot the gate-dependent PL response for the 100 and 150\,nm finger gate widths to assess the generality of the confinement effect. Fig.~\ref{fig-sup:PL_otherfingers}\bfA\, shows PL spectra as a function of $\vbg$ at fixed $\vtg$=0 V, showing the known dependence of the interface state with respect to doping. When the top-gate voltage is set to $\vtg$= -5 V (Fig.~\ref{fig-sup:PL_otherfingers}\bfB), a new spectral feature emerges near $\vbg$=3 V. This feature appears in a voltage window where the surrounding interface states have already undergone a large blueshift, yet the confined X$_{\mathrm{trap}}$ excitons remain neutral, corresponding to the previously mentioned n-i-n regime. The confined states exhibit a qualitatively similar doping dependence to the unconfined excitons but in a different gate range, providing direct evidence of lateral confinement. Figs.~\ref{fig-sup:PL_otherfingers}\bfC\, and \bfD\, present the analogous data for the 150\,nm wide finger gate, showing the same phenomenology and supporting the interpretation that confinement is a robust feature across different finger widths.

\subsection{Additional data for the power measurements}

\begin{figure}[!htbp]
\centering
\includegraphics[width=\linewidth]{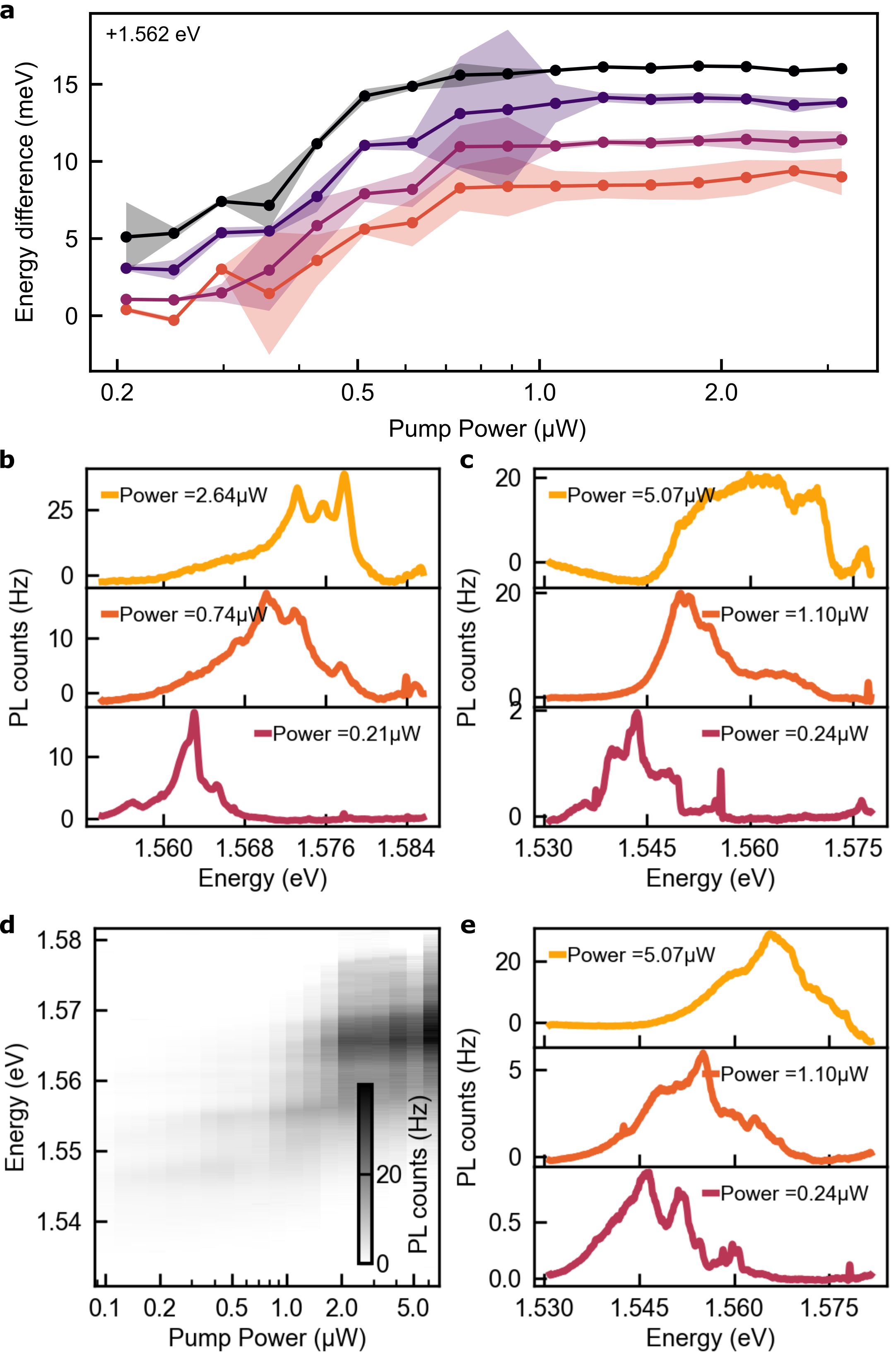}
\caption{\textbf{Additional power dependent data}. \textbf{(a)} Fitted center wavelengths of the individual emission peaks of the confined X$_{\mathrm{trap}}$ excitons below the 50 nm finger, as a function of the excitation laser power. The fitting error is shown as the shaded region around each line. \textbf{(b)} Raw PL spectra of the confined X$_{\mathrm{LI}}$ excitons below the 50\,nm finger at different powers. The states exhibit a common blueshift with power as well as a redistribution of the emission intensity between the different sub-bands. \textbf{(c)} Raw PL spectra of the confined X$_{\mathrm{LI}}$ excitons below the 100\,nm finger at different powers. The states exhibit a common blueshift with power as well as a redistribution of the emission intensity between the different sub-bands. \textbf{(d)} Energy dependence of the confined X$_{\mathrm{LI}}$ states below the 150\,nm finger with power. The state bundle experiences a 10 meV blue shift that stops after the excitation power reaches $\approx1$\,\textmu W. \textbf{(e)} Raw PL spectra of the confined X$_{\mathrm{LI}}$ excitons below the 150\,nm finger at different powers.}
\label{fig-sup:power}
\end{figure}

To further characterize the nonlinear optical response of the confined X$_{\mathrm{trap}}$ states, we performed power-dependent PL measurements on all three finger gate widths. For the 50\,nm finger (Figs.~\ref{fig-sup:four}\bfA-\bfB), we tracked the fitted center energies of the individual emission peaks as a function of excitation power, with fitting uncertainties shown as shaded bands. The fitting procedure is the same as the one used previously on Ref.\cite{Vandoolaeghe2025}. The peaks exhibit a collective blueshift with increasing power, accompanied by a redistribution of emission intensity among the different sub-bands shown in the raw spectra with increasing power, suggesting a power-dependent repopulation of the confined states driven by dipolar interactions between interfacial excitons. For the 100\,nm finger (Fig.~\ref{fig-sup:power}\bfC), the corresponding raw spectra confirm this behavior. Figs.~\ref{fig-sup:power}\bfD\, and \bfE\, present the equivalent dataset for the 150\,nm wide finger gate, showing a qualitatively similar power dependence, reinforcing the conclusion that the observed blueshift and band redistribution are intrinsic to the confined interfacial exciton system and scale consistently across confinement geometries of different widths.

\subsection{Additional data for the photonic time correlation}

\begin{figure}[!htbp]
\centering
\includegraphics[width=\linewidth]{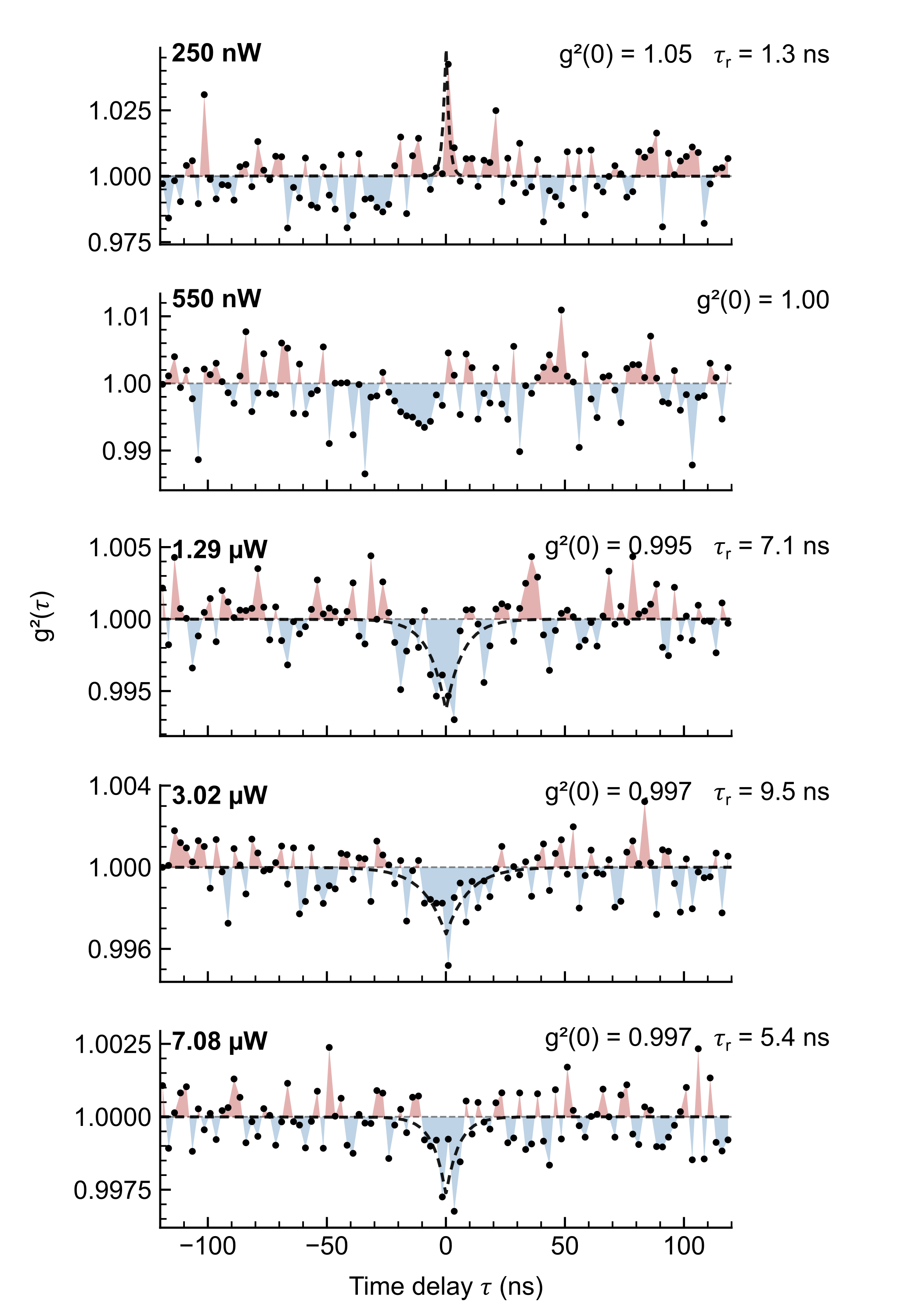}
\caption{\textbf{$g^{(2)}_\text{ph}(\tau)$ for the 150 nm finger.} As with the other two fingers, a transition from bunching at low powers to antibunching at high powers is observed, albeit with a reduced contrast given the system size. }
\label{fig-sup:four}
\end{figure}

We present here the measured correlation data coming from the X$_{\mathrm{trap}}$ excitons below the 150 nm finger in Fig.~\ref{fig-sup:four}. It exhibits the same trend as what is shown in the main text for the thinner fingers, with a reduced contrast as expected from Eq 1 of the main text.

\subsection{Time-correlated single photon counting}

\begin{figure}[!htbp]
\centering
\includegraphics[width=\linewidth]{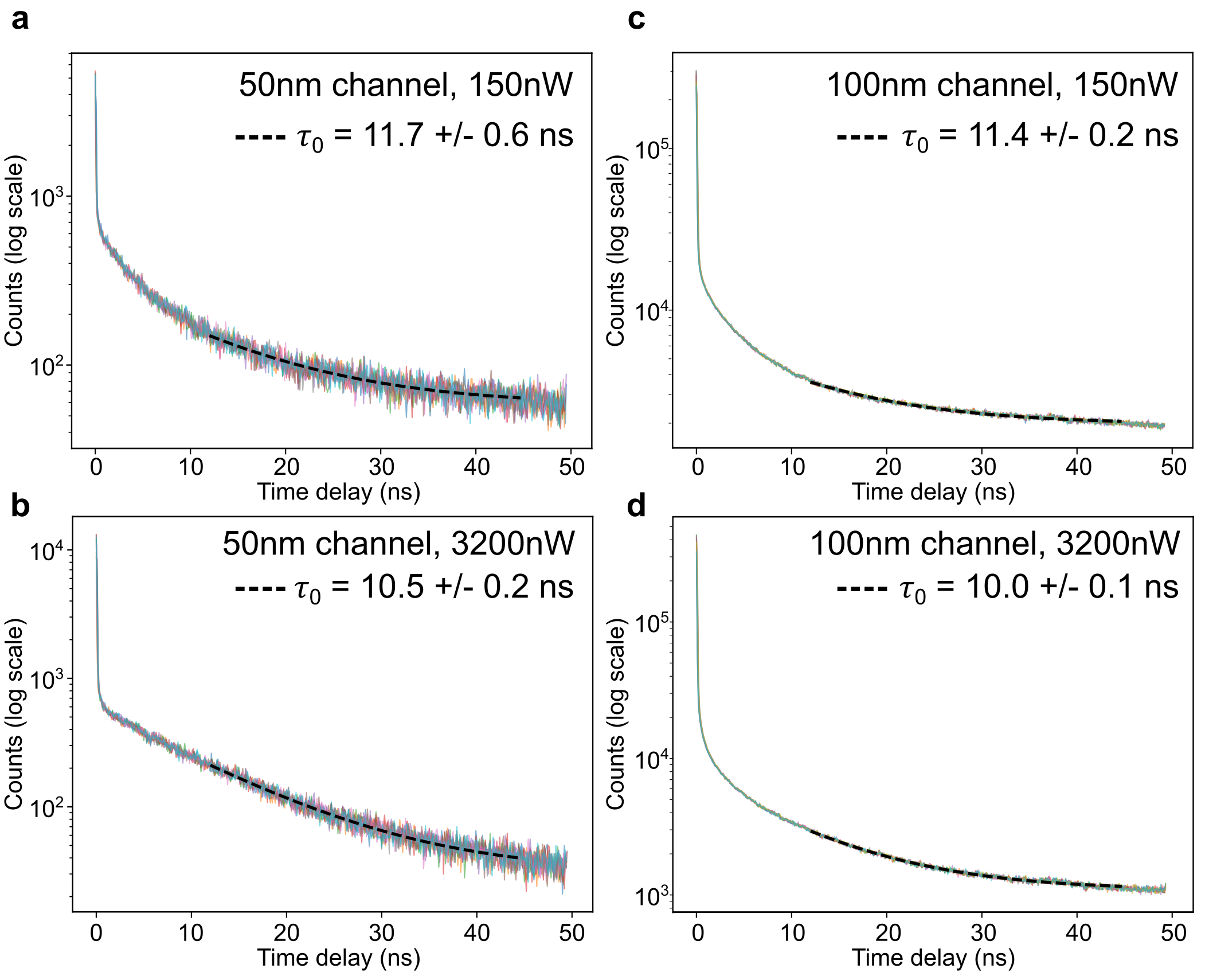}
\caption{\textbf{Time-correlated single-photon counting histograms.}    (\textbf{a}) Fluorescent decay trace for the excitons in the 50\,nm trap under 150\,nW of excitation power. A single-exponential tail fit gives a characteristic lifetime of $\tau_0 \approx$ 11.7 ns. (\textbf{b}) Same data for 3200\,nW of excitation power, yielding $\tau_0 \approx$ 10.5 ns. Panels (\textbf{c}) and (\textbf{d}) show the fluorescent decay traces for the 100\,nm trap, with similar extracted $\tau_0 \approx$ 11.4\,ns at low power and  $\tau_0 \approx$ 10\,ns at high power.}
\label{fig-sup:five}
\end{figure}

We present in Fig.\ref{fig-sup:five} representative fluorescent lifetime measurements of X$_{\mathrm{trap}}$ for different channel length and pump power, as obtained by time-correlated single photon counting (TCSPC). The fitted lifetime is obtained by single-exponential tail-fit on the decay histograms. The zero-delay spike in the histogram is due to residual MoSe$_2$ trion  emission tails, collected by the SNSPDs through the long-pass spectral filters. Note that the pulsed excitation regime used here differs from the continuous wave (CW) excitation regime presented throughout the manuscript. Taking $\tau_0\sim10\,$ns, a pump repetition period $T=50\,$ns, and assuming linearity, the pulsed excitation is expected to create a peak exciton population $\sim5\,$x the steady-state population of the CW case for equal average pump powers.

\subsection{Ruling out alternative mechanisms for the observed photon statistics}

The bunching-to-antibunching crossover reported in the main text is interpreted as the optical fingerprint of a many-body phase crossover in the trapped 1D dipolar exciton gas. Here we examine three single-particle or kinetic alternatives that could in principle produce similar signatures without invoking collective matter correlations, and show that each is inconsistent with the data.

\subsubsection{Ensemble of localized defects}

The most direct extrinsic origin for sub-Poissonian photon statistics is emission from an ensemble of $N$ independent, two-level defect emitters. Such an ensemble is structurally bounded by $g^{(2)}_{\rm ph}(0) \le 1 - 1/N$ at all drive strengths, irrespective of spectral inhomogeneity, individual saturation behavior, or spatial arrangement. In particular, no such ensemble can produce $g^{(2)}_{\rm ph}(0) > 1$. The observation of strong bunching $g^{(2)}_{\rm ph}(0) \approx 2$ at low power in the $L = 50\,\mathrm{nm}$ trap (Fig.~4 \bfA,\bfC) therefore falsifies the defect-ensemble hypothesis by itself: bunching at this level requires bosonic interference within a single coherence cell, which cannot be reproduced by any number or arrangement of independent two-level defects.

This argument extends to hybrid models --- bright defects superposed on an interacting background, defect-bound trions, or defect-hybridized tail states --- which are excluded by two further features of the data. (i) The bunching contrast scales with trap size as $g^{(2)}_{\rm ph}(0) \approx 1 + \xi_{\rm coh}/L$ (Eq.~1), saturating near 2 at $L = 50\,\mathrm{nm}$ and decreasing to $\approx 1.5$ at $L = 100\,\mathrm{nm}$, yielding a matter coherence length $\xi_{\rm coh} \approx 50\,\mathrm{nm}$. A defect-dominated model predicts the opposite scaling: larger traps host more independent emitters, pushing $g^{(2)}_{\rm ph}(0)$ from below toward unity. (ii) The $\sim 12\,\mathrm{meV}$ density-dependent blueshift correlates one-to-one with PL intensity rather than with gate-set carrier density, and is consistent with a collective Hartree shift of an interacting gas; defect emission produces narrow, fixed-energy lines, and environment-mediated shifts (Stark, screening) would track gate configuration rather than trapped-state intensity. Together these features rule out defect emission --- pure or hybridized with an interacting background --- as the origin of the bunching-to-antibunching evolution.

\subsubsection{Disorder-induced sub-traps and non-interacting few-level filling}

Two related single-particle alternatives must be considered: residual disorder fragmenting the trap into localized sub-traps, or sequential filling of the non-interacting transverse subband ladder. Each is ruled out by an independent feature of the data.

\textbf{Disorder-induced sub-traps.} Three independent considerations exclude disorder fragmentation. First, the mesoscopic gating geometry itself suppresses the role of disorder. Conventional diffraction-limited optical experiments sample a $\sim 1\,\mu\mathrm{m}$ region and can therefore average over many disorder configurations. Our gate-defined traps ($L = 50$--$150\,\mathrm{nm}$) are much smaller than the diffraction limit, therefore the role of disorder averaging is significantly reduced.

Second, the low-power photon statistics confirm this empirically. The chaotic single-mode value $g^{(2)}_{\rm ph}(0) \approx 2$ in the $L = 50\,\mathrm{nm}$ trap requires emission from a single coherent matter cell with $\xi_{\rm coh} \sim L$; multiple sub-traps within $50\,\mathrm{nm}$ would suppress the bunching below 2 by a factor $\sim \xi_{\rm coh}^{\rm disorder}/L$. The bunching saturating at the single-mode value in the smallest trap and decreasing to $\approx 1.5$ at $L = 100\,\mathrm{nm}$ directly demonstrates a single coherent mode rather than disorder fragments.

Third, in the regime where antibunching actually develops, the interaction-induced chemical potential reaches $\mu \sim 12\,\mathrm{meV}$ --- far exceeding the few-meV disorder energy scale typical of these heterostructures \cite{Cadiz2017, Raja2019}. Residual disorder is therefore washed out by interactions precisely where antibunching is observed: the strongly correlated phase delocalizes excitons over the full trap regardless of the low-density disorder landscape.

\textbf{Non-interacting few-level filling.} The clearest signature distinguishing many-body interactions from non-interacting level filling lies in the joint evolution of the transverse subband energies and the brightness distribution among them (Fig.3 \bfA). A non-interacting picture predicts fixed subband energies, with filling redistributing brightness up the ladder while each subband emits at its own fixed energy. The observation is the opposite. As power increases, \emph{all} transverse subbands shift smoothly and together toward higher energies before saturating, while the brightness weight transfers sequentially up the ladder. The energy shift is therefore \emph{collective}, \emph{smooth}, and \emph{decoupled from the brightness transfer}. This is the canonical Hartree signature of a mean-field interaction $\mu(n)$ felt by every exciton in the trap, which cannot arise in any non-interacting few-level model.

An environment-mediated alternative --- collective shifts from photogenerated-carrier-induced screening or Stark effects --- is also ruled out, since such effects would track the gating-set carrier density rather than the trapped-exciton population, whereas the observed blueshift correlates one-to-one with the trapped-state PL intensity and saturates at the same $P_{\rm sat}$ (Fig.3 \bfB\,,\bfC\,).

The total blueshift of $\sim 12\,\mathrm{meV}$, several times larger than $\hbar\omega_y \approx 3$--$5\,\mathrm{meV}$, sets the quantitative scale of the dipolar interaction energy at $n \approx 10^2\,\mu\mathrm{m}^{-1}$ and is consistent with the estimated dipolar coupling $C_{dd}/d^3$ (see Sec.~S3 C).

\subsubsection{Auger recombination}

Auger recombination (exciton--exciton annihilation) is the most common kinetic route to sub-Poissonian photon statistics: pairwise non-radiative removal of excitons suppresses two-photon coincidences without invoking any equilibrium correlation hole. We rule it out on both dynamical and energetic grounds.

\textbf{Dynamical signature.} The defining signature of Auger is a density-dependent shortening of the exciton lifetime via the additional decay channel $\Gamma_{\rm Auger} \propto C_A n$ \cite{Moody2016}. 
Time-correlated single-photon counting experiments shows instead a relatively constant lifetime, with $\tau_0 = 11\pm2\,$ns across different channels and pump intensities.
%Time-resolved photoluminescence shows the opposite: $\tau_0$ \emph{increases} from $\sim\!5$ to $\sim\!15\,\mathrm{ns}$ across the crossover. 
Moreover, the correlation recovery time $\tau_r$ --- which reflects the relaxation of density fluctuations and need not coincide with $\tau_0$ --- rises sharply at high power and exceeds the radiative lifetime, a dynamical signature of collective rather than single-particle relaxation. Auger predicts neither trend; it would shorten both $\tau_0$ and $\tau_r$ with density.

\textbf{Energetic signature.} Auger does not shift the emission energy of surviving excitons; it removes pairs and leaves the remainder spectroscopically unchanged. The $\sim 12\,\mathrm{meV}$ density-dependent blueshift, saturating in lockstep with the intensity, is therefore inconsistent with an Auger-dominated scenario and requires interaction-induced stiffening of $\mu$.

\textbf{Consistency with the dipolar picture.} The above arguments rule out Auger without invoking any specific model of the many-body state. As a consistency check, the 1D dipolar Bose gas in the strongly correlated regime develops a deep short-range correlation hole --- $g^{(2)}_{\rm mat}(r=0) \to 0$ exactly in the quasi-crystal (fermionized) limit, and strongly suppressed but finite across our accessed parameter range. The same short-range repulsion that produces this hole kinetically suppresses the close-approach overlap required for Auger, offering an a posteriori explanation for why Auger is absent here even at densities where it would dominate intralayer TMD excitons. The orders of magnitude reduction of Auger annihilation rate in interlayer dipolar excitons, due to repulsion, has been observed in Ref.\cite{Cai2024}, which further supports our view that Auger is negligible in our system.

\section{Theoretical aspects of Photon Correlation Microscopy of quantum matter}
\subsection{Photon correlation mapping equation}
Here we derive the photon correlation microscopy identity [Eq.~(1) of the main text], which maps the density correlations of a many-body system onto the photon-pair statistics of its emitted light, and we state the conditions of validity and the limiting forms used to interpret the data.
 
\subsubsection{Matter identity}
 
Consider a many-body system described by field operators $\hat{\psi}(x)$ with
density $\hat{n}(x) = \hat{\psi}^\dagger(x)\hat{\psi}(x)$ and total number
$\hat{N} = \int_L \hat{n}(x)\,dx$ in a region of extent $L$. The normalized
spatial pair correlator is
\begin{equation}
g^{(2)}_{\rm mat}(x,x') =
\frac{\langle \hat{\psi}^\dagger(x)\hat{\psi}^\dagger(x')
\hat{\psi}(x')\hat{\psi}(x)\rangle}
{\langle \hat{n}(x)\rangle\langle \hat{n}(x')\rangle}.
\label{eq:g2mat_def}
\end{equation}
For bosons, the commutator $[\hat{\psi}(x),\hat{\psi}^\dagger(x')] =
\delta(x-x')$ gives the normal-ordering rearrangement
\begin{equation}
\hat{\psi}^\dagger(x)\hat{\psi}^\dagger(x')\hat{\psi}(x')\hat{\psi}(x)
= \hat{n}(x)\hat{n}(x') - \delta(x-x')\,\hat{n}(x).
\end{equation}
Integrating both coordinates over the region $L$ gives the exact identity

\begin{widetext}
\begin{equation}
    \iint_L \langle \hat{\psi}^\dagger(x)\hat{\psi}^\dagger(x')
    \hat{\psi}(x')\hat{\psi}(x)\rangle\,dx\,dx'
    = \iint_L \big[\langle \hat{n}(x)\hat{n}(x')\rangle
    - \delta(x-x')\langle \hat{n}(x)\rangle\big]\,dx\,dx' \\
    = \langle \hat{N}^2\rangle - \langle \hat{N}\rangle
    = \langle \hat{N}(\hat{N}-1)\rangle.
    \label{eq:factorial}
\end{equation}
\end{widetext}

The $-\langle\hat{N}\rangle$ contribution, and hence the Poisson floor that
appears below, originates entirely from the $\delta(x-x')$ self-correlation.
For fermions the anticommutator changes the sign of the contact term in the
field-operator rearrangement, but the integrated result is unchanged:
$\langle\hat{N}(\hat{N}-1)\rangle = \langle\hat{N}^2\rangle -
\langle\hat{N}\rangle$ is the second factorial moment of the number
distribution $p(N)$ and is independent of exchange statistics. Equation
\eqref{eq:factorial} therefore holds for bosons, fermions, and anyons alike;
only the short-range form of $g^{(2)}_{\rm mat}$ differs (e.g.\ an exchange
hole for fermions).
 
Using the definition \eqref{eq:g2mat_def} to write
$\langle\hat{\psi}^\dagger(x)\hat{\psi}^\dagger(x')\hat{\psi}(x')
\hat{\psi}(x)\rangle = \langle n(x)\rangle\langle n(x')\rangle\,
g^{(2)}_{\rm mat}(x,x')$, Eq.~\eqref{eq:factorial} becomes the exact,
general matter identity
\begin{align}
\frac{\langle \hat{N}(\hat{N}-1)\rangle}{\langle \hat{N}\rangle^2} &
= \frac{1}{\langle N\rangle^2}\iint_L
\langle n(x)\rangle\langle n(x')\rangle\,
g^{(2)}_{\rm mat}(x,x')\,dx\,dx',
% &\langle N\rangle \equiv \int_L \langle n(x)\rangle\,dx,
\label{eq:matter_general}
\end{align}
valid for any density profile and any exchange statistics.
 
\subsubsection{Homogeneous reduction and the finite-window kernel}
 
In the main text we quote the homogeneous form of
Eq.~\eqref{eq:matter_general}. For a translation-invariant system,
$\langle n(x)\rangle = \bar n$ is uniform and $g^{(2)}_{\rm mat}(x,x')$
depends only on the separation $r = x-x'$. The constant density factors then
cancel against $\langle N\rangle^2 = (\bar n L)^2$, giving
\begin{align}
\frac{1}{\langle N\rangle^2}\iint_L
\langle n(x)\rangle\langle n(x')\rangle\,
g^{(2)}_{\rm mat}(x,x')\,dx\,dx' \\
= \frac{1}{L^2}\iint_L g^{(2)}_{\rm mat}(x,x')\,dx\,dx'.
\end{align}
Changing variables to $r = x-x'$ and $R = (x+x')/2$ (unit Jacobian), the
integrand depends only on $r$, while the range of $R$ available at fixed $r$
inside the window $[0,L]$ has length $L-|r|$ for $|r|\le L$ and zero
otherwise. Performing the $R$ integral yields the finite-window
(triangular) kernel
\begin{equation}
\frac{1}{L^2}\iint_L g^{(2)}_{\rm mat}(x,x')\,dx\,dx'
= \frac{1}{L}\int_{-L}^{L} g^{(2)}_{\rm mat}(r)
\left(1-\frac{|r|}{L}\right)dr.
\label{eq:triangular}
\end{equation}
The weight $1-|r|/L$ is the autocorrelation of the sampling window with
itself: the number of pairs of points in $[0,L]$ separated by $r$ falls
linearly as $|r|$ grows. It is a finite-size geometric factor, not a
physical correlation effect, and reduces to unity for $|r|\ll L$. Two
checks: for an uncorrelated system ($g^{(2)}_{\rm mat}=1$) the right-hand
side evaluates to $1$, as required; and for a correlation length
$\xi\ll L$, the kernel is $\approx 1$ over the support of
$g^{(2)}_{\rm mat}(r)-1$ and Eq.~\eqref{eq:triangular} reduces to
$1 + L^{-1}\!\int[g^{(2)}_{\rm mat}(r)-1]\,dr$, the $\xi/L$ dilution that
underlies the mesoscopic-sampling argument and the $1+\xi_{\rm coh}/L$
scaling of the bunching amplitude in the main text. When $\xi\sim L$ the
kernel cannot be dropped, and its falloff is precisely what reduces the
measured bunching below the ideal value $g^{(2)}_{\rm mat}(0)$ in the
larger traps. \\

\subsubsection{Photon mapping}
 
The light field inherits this structure when the emission is incoherent.
For spontaneous emission in which each particle radiates independently at
rate $\Gamma$, the instantaneous intensity is $I(t)\propto\hat{N}(t)\Gamma$
and the photon detection normal-ordering inherits the matter
normal-ordering. The zero-delay second-order coherence measured in a
Hanbury Brown--Twiss interferometer is then
\begin{equation}
g^{(2)}_{\rm ph}(0) = \frac{\langle :\!I^2\!:\rangle}{\langle I\rangle^2}
= \frac{\langle \hat{N}(\hat{N}-1)\rangle}{\langle \hat{N}\rangle^2},
\end{equation}
where $:\,:$ denotes normal ordering. This identifies the photon observable
with the matter quantity in Eq.~\eqref{eq:matter_general}. The only physical
input is the proportionality condition: that the emitted flux tracks the
instantaneous local matter density, with incoherent emission so that no
first-order coherence cross term contributes.
 
\subsubsection{Compressibility identity}
 
In quasi-equilibrium the number fluctuations are fixed by a thermodynamic
susceptibility. The grand-canonical fluctuation--dissipation relation reads
\begin{align}
\mathrm{Var}(N) =& n\,k_B T\,\langle N\rangle\,\kappa_T
\nonumber \\
\qquad
\kappa_T \equiv& \frac{1}{n^2}\left(\frac{\partial n}{\partial \mu}\right)_T,
\label{eq:fdt}
\end{align}
with $n=\langle N\rangle/L$. Here $\kappa_T$ is the isothermal
compressibility written in its density (number-response) form; this is the
same definition used in the main text and in the Fig.~3 caption. It is
convenient to separate the connected part of the correlator. Writing
$g^{(2)}_{\rm mat} = 1 + [\,g^{(2)}_{\rm mat}-1\,]$ in
Eq.~\eqref{eq:matter_general}, the constant term integrates to unity and
\begin{align}
g^{(2)}_{\rm ph}(0) =& 1 +
\frac{1}{\langle N\rangle^2}\iint_L
\langle n(x)\rangle\langle n(x')\rangle\,
\big[\,g^{(2)}_{\rm mat}(x,x')-1\,\big]\,dx\,dx' \\
=& 1 + \frac{\mathrm{Var}(N)-\langle N\rangle}{\langle N\rangle^2}.
\label{eq:connected}
\end{align}
Equation~\eqref{eq:connected} makes explicit that the sign of
$g^{(2)}_{\rm ph}(0)-1$ is set by the integrated connected correlator:
bunching for a positive (correlation-bump) integrand, antibunching for a
negative (correlation-hole) one. Substituting the
fluctuation--dissipation relation \eqref{eq:fdt} for $\mathrm{Var}(N)$
gives the third member of the master identity, and collecting all three:
\begin{eqnarray}
g^{(2)}_{\rm ph}(0) =&
\underbrace{\frac{\langle N(N-1)\rangle}{\langle N\rangle^2}}_{\rm Fluctuations} \\
=& \underbrace{\frac{1}{L^2}\iint_L g^{(2)}_{\rm mat}(x,x')\,dx\,dx'}_{\rm Correlations} \\
=& \underbrace{1 + \frac{k_B T \kappa_T}{L} - \frac{1}{\langle N\rangle}}_{\rm Compressibility},
\label{eq:master}
\end{eqnarray}
where the second member is written in its homogeneous form
[Eq.~\eqref{eq:triangular}, with the window kernel implicit]; the general
inhomogeneous expression is Eq.~\eqref{eq:matter_general}.
 
\subsubsection{Conditions of validity and limiting forms}
 
The identities in Eq.~\eqref{eq:master} have a clear hierarchy of
assumptions.
 
\paragraph{First identity (fluctuations).} The equality
$g^{(2)}_{\rm ph}(0) = \langle N(N-1)\rangle/\langle N\rangle^2$ is exact.
It requires only the proportionality condition (incoherent,
density-tracking emission) and holds for any steady state with
$\langle N\rangle\ge 1$, independent of exchange statistics. In particular,
$g^{(2)}_{\rm ph}(0)<1$ is an exact, model-independent signature of
sub-Poissonian number fluctuations, $\mathrm{Var}(N)<\langle N\rangle$.
 
\paragraph{Second identity (correlations).} The general form
Eq.~\eqref{eq:matter_general} is likewise exact for any density profile.
The homogeneous form quoted in the main text additionally assumes
translation invariance over the sampling window; the finite-window kernel
$1-|r|/L$ [Eq.~\eqref{eq:triangular}] is exact within that assumption, and
the residual inhomogeneity of the flat-bottomed trap (Fig.~2 \bfC) sets the
small correction to it.
 
\paragraph{Third identity (compressibility).} This additionally invokes
the grand-canonical fluctuation--dissipation relation \eqref{eq:fdt}, and
is therefore a quasi-equilibrium statement. The compressibility $\kappa_T$
is precisely defined as an intensive susceptibility only for
$\langle N\rangle\gg 1$; the relation is asymptotically valid in the
mesoscopic regime and degenerates smoothly toward the few-particle limit.

\subsubsection{Limiting regimes}
\noindent
\textbf{1. Thermal Bose gas.} For a non-degenerate 1D Bose gas ($n\lambda_T \lesssim 1$), the pair correlator factorizes as $g^{(2)}_{\rm mat}(r) = 1 + |g^{(1)}(r)|^2$ with a Gaussian first-order coherence $|g^{(1)}(r)|^2 = e^{-r^2/\lambda_T^2}$ on the scale of the thermal de Broglie wavelength $\lambda_T = \hbar\sqrt{2\pi/(m k_B T)}$. Integrating over a trap of length $L$:

\begin{equation}
g^{(2)}_{\rm ph}(0) = 1 + \frac{1}{L}\int^{-L}_L e^{-r^2/{\lambda_T}^2}( 1- |r|/L) dr
\end{equation}
which interpolates between $g^{(2)}_{\rm ph}(0) \to 2$ for $\lambda_T \gg L$ (single-mode chaotic limit) and $g^{(2)}_{\rm ph}(0) \to 1 + \sqrt{\pi}\,\lambda_T/L$ for $\lambda_T \ll L$ (multimode bulk limit). The bulk form can equivalently be written as $g^{(2)}_{\rm ph}(0) \to 1 + 1/M$ where $M \sim L/\lambda_T$ is the effective number of coherence cells sampled. \\

\noindent
\textbf{2. Quantum-degenerate regime.} As $n\lambda_T \gtrsim 1$, number fluctuations approach the Poissonian value $\mathrm{Var}(N) \to \langle N\rangle$, the spatial pair correlator approaches unity across the trap, and $g^{(2)}_{\rm ph}(0) \to 1$. \\

\noindent
\textbf{3. Strongly correlated incompressible regime.} As repulsive interactions stiffen the chemical potential, $\kappa_T \to 0$ and the system develops a correlation hole at short range, $g^{(2)}_{\rm mat}(0) \to 0$. The master identity then yields the Poisson floor. \\

\begin{equation}
g^{(2)}_{\rm ph}(0) \to 1 - \frac{1}{\langle N\rangle},
\end{equation}
attained only in the perfectly number-stabilized limit. For the specific case of contact-interacting bosons in the Tonks--Girardeau (fermionized) limit, the matter pair correlator takes the closed form $g^{(2)}_{\rm mat}(r) = 1 - \mathrm{sinc}^2(\pi n r)$, where $\mathrm{sinc}(x)=\mathrm{sin}(x)/x$ is the cardinal sine function; for dipolar interactions, the strongly correlated regime develops a qualitatively similar correlation hole with oscillations at the inter-particle spacing $1/n$, but with a different functional form. \\

\noindent
\textbf{4. Classical chaotic emission.} A complementary single-mode result arises classically: $N$ independent emitters with random phases produce a field $E = \sum_j a_j e^{i\phi_j}$ that follows a complex Gaussian distribution by the central limit theorem, yielding exponentially distributed intensity and $g^{(2)}_{\rm ph}(0) = 2$. This is the classical chaotic limit, identical in value to the single-mode thermal Bose result but arising from random-phase wave interference rather than from matter number fluctuations. The master identity above applies to the latter; we note this distinction because it clarifies that $g^{(2)}_{\rm ph}(0) = 2$ at low power in our smallest trap reflects coherent bosonic statistics in a single matter mode (an interference effect of identical particles), not classical chaotic phase of distinguishable emitters.

\begin{table*}[ht!]
\centering
\renewcommand{\arraystretch}{1.6}
\setlength{\tabcolsep}{8pt}
\begin{tabular}{lcccc}
\hline
\textbf{Regime} & $\mathbf{\mathrm{Var}(N)}$ & $\boldsymbol{\kappa_T}$ & $\mathbf{g^{(2)}_{\rm mat}(r)}$ & $\mathbf{g^{(2)}_{\rm ph}(0)}$ \\
\hline
Classical chaotic single-mode (wave interference) & $\langle N\rangle$ & $\dfrac{1}{n k_B T}$ & $1$ & $2$ \\
Thermal Bose gas & $\langle N\rangle(1+\langle N\rangle/M)$ & $\dfrac{1+\langle N\rangle/M}{n k_B T}$ & $1 + e^{-r^2/\lambda_T^2}$ & 1 + 1/M \\
Quantum-degenerate ($n\lambda_T \gtrsim 1$) & $\langle N\rangle$ & $\dfrac{1}{n k_B T}$ & $\to 1$ & $\to 1$ \\
Incompressible / fermionized & $\to 0$ & $\to 0$ & oscillatory, $g^{(2)}_{\rm mat}(0) \to 0$ & $\to 1 - 1/\langle N\rangle$ \\
\hline
\end{tabular}
\caption{Limiting regimes of the master identity. The thermal Bose entry interpolates between single-mode chaotic bunching ($L \ll \lambda_T$, $M \sim 1$) and multimode bulk behavior ($L \gg \lambda_T$, $M \sim L/\lambda_T$) through Eq.~\ref{eq:g2_thermal}. The first three rows are described by the master identity; the fourth row (classical chaotic single-mode) gives $g^{(2)}_{\rm ph}(0) = 2$ via a wave-interference mechanism outside the scope of the master identity, included for comparison.}
\end{table*}

\subsection{Effective exciton temperature}

While the cryostat operates at $T_{\rm bath} \approx 30\,\mathrm{mK}$, the effective temperature $T_{\rm eff}$ of the optically driven exciton gas is set by phonon-mediated cooling of the photogenerated carriers and may be substantially higher. The low-power bunching contrast provides a direct, in-situ measurement of $T_{\rm eff}$ via the matter coherence length $\xi_{\rm coh}$.

In the thermal regime ($n \lambda_T \lesssim 1$), the matter pair correlator of a 1D Bose gas is $g^{(2)}_{\rm mat}(r) = 1 + |g^{(1)}(r)|^2$ with $|g^{(1)}(r)|^2 = e^{-r^2/\xi_{\rm coh}^2}$ and $\xi_{\rm coh} \approx \lambda_T = \hbar\sqrt{2\pi/(m_X k_B T_{\rm eff})}$. Substituting into Eq.1 and integrating over the trap yields

\begin{equation}
g^{(2)}_{\rm ph}(0) = 1 + \frac{\sqrt{\pi}\,\xi_{\rm coh}}{L}\,\mathrm{erf}\!\left(\frac{L}{\xi_{\rm coh}}\right) - \frac{\xi_{\rm coh}^2}{L^2}\!\left(1 - e^{-L^2/\xi_{\rm coh}^2}\right),
\label{eq:g2_thermal}
\end{equation}
which interpolates between $g^{(2)}_{\rm ph}(0) \to 2$ for $\xi_{\rm coh} \gg L$ and $g^{(2)}_{\rm ph}(0) \to 1 + \sqrt{\pi}\,\xi_{\rm coh}/L$ for $\xi_{\rm coh} \ll L$. Inverting Eq.~\ref{eq:g2_thermal} with the measured $g^{(2)}_{\rm ph}(0) \approx 1.5$ at $L = 100\,\mathrm{nm}$ pins $\xi_{\rm coh} \approx 50\,\mathrm{nm}$; the saturated value $g^{(2)}_{\rm ph}(0) \approx 2$ in the $50\,\mathrm{nm}$ trap is consistent with this but provides only a lower bound. Using $m_X = 1.3\,m_e$ (Ref.~\cite{Vandoolaeghe2025}),

\begin{equation}
T_{\rm eff} = \frac{2\pi\hbar^2}{m_X k_B \xi_{\rm coh}^2} \approx 1.7\,\mathrm{K}.
\end{equation}

The dominant uncertainties --- the assumed shape of $g^{(1)}(r)$ ($\sim 10\%$) and quantum-degenerate corrections at the $n\lambda_T \sim 1$ crossover --- place the true value in the range $1$--$2\,\mathrm{K}$.%, with the estimate above acting as an upper bound.

This $T_{\rm eff}$ is consistent with three independent constraints. (i) It exceeds the cryostat temperature by two orders of magnitude, as expected for an optically driven exciton gas cooling toward the lattice on the picosecond exciton--phonon timescale. (ii) The thermal energy $k_B T_{\rm eff} \approx 0.15\,\mathrm{meV}$ is far below the trap depth ($\sim 20-40\,\mathrm{meV}$) and the transverse subband spacing ($\hbar\omega_y \approx 3$--$5\,\mathrm{meV}$), consistent with the lowest-subband dominance at low power (Fig.3 \bfA). (iii) At high power, $\mu/k_B T_{\rm eff} \approx 80 \gg 1$, placing the system deep in the interaction-dominated quantum-degenerate regime. The thermometry --- extracted from the photon-correlation data itself --- validates the use of equilibrium thermodynamic observables in interpreting the bunching-to-antibunching crossover.

\subsection{Thermalization and interaction timescales}

The validity of Eq.~(1) requires that the matter reaches a quasi-equilibrium state between successive emission events:
\begin{equation}
    \tau_\text{int},\ \tau_\text{th} \;\ll\; \Gamma^{-1},
    \tag{S1}
\end{equation}
where $\tau_\text{int}$ is the timescale of dipolar interactions in the gas, $\tau_\text{th}$ is the phonon-mediated thermalization time, and $\Gamma^{-1}$ is the radiative lifetime.  This appendix derives each timescale from independent inputs and verifies the hierarchy across the operating range.

\subsubsection*{Interaction timescale $\tau_\text{int}$}

The natural timescale for matter-side dynamics is set by the inverse of the dipolar mean-field chemical potential, $\tau_\text{int} = \hbar/\mu_\text{int}(n)$, where $\mu_\text{int}(n)$ is the interaction contribution to the chemical potential of the 1D dipolar Bose gas. To Hartree order with quantum-pressure correction \cite{Astrakharchik2008,Arkhipov2005},
\begin{equation}
    \mu_\text{int}(n) \;\approx\; 4\, C_{dd}\, n^3 \;+\; \pi^2 \frac{\hbar^2}{2 m^*}\, n^2,
    \tag{S2}
\end{equation}
where the first (cubic) term is the Hartree dipolar contribution from a 1D gas of point dipoles at mean separation $1/n$, and the second (quadratic) term is the Tonks--Girardeau-like Fermi pressure that becomes important as $n r_0 \gtrsim 1$.  The two parameters are
\begin{equation}
    C_{dd} = \frac{e^2 d^2}{4\pi\epsilon_0 \epsilon_r}, \quad
    r_0 = \frac{m^* C_{dd}}{\hbar^2},
    \tag{S3}
\end{equation}
with $d = 2.2\,\text{nm}$ the interface dipole length, $m^* = 1.3\,m_e$ the in-plane exciton mass, and $\epsilon_r$ the in-plane effective dielectric constant of the hBN-encapsulated heterostructure.  Plugging in numbers, $C_{dd} \approx 1.4 \times 10^3$ meV$\cdot$nm$^3$, $\hbar^2/2m^* \approx 47.6$ meV$\cdot$nm$^2$, and $r_0 \approx 14.6$ nm.

The interaction time follows directly:
\begin{equation}
    \tau_\text{int}(n) \;=\; \hbar / \mu_\text{int}(n).
    \tag{S4}
\end{equation}

% \begin{center}
% \begin{tabular}{cccc}
% \hline
% $n$ (/nm)  & $n r_0$  & $\mu_\text{int}$ (meV)  & $\tau_\text{int}$ (ps)  \\
% \hline
% 0.02 & 0.29 & 0.21 & 3.1 \\
% 0.05 & 0.73 & 1.9  & 0.35 \\
% 0.10 & 1.46 & 10.3 & 0.064 \\
% 0.13 & 1.90 & 17.1 & 0.039 \\
% \hline
% \end{tabular}
% \end{center}

At the lowest experimentally accessed densities ($n r_0 \ll 1$), $\tau_\text{int}$ is of order a few ps; at saturation densities ($n r_0 \sim 1.5$, $\mu_\text{int} \sim 10\text{--}15$ meV consistent with the observed blueshift), $\tau_\text{int}$ drops below 100 fs. Across the bunching-to-antibunching crossover, $\tau_\text{int}$ remains in the sub-picosecond to few-picosecond range.

\subsubsection*{Thermalization timescale}

The validity of Eq.~(1) as a quasi-equilibrium relation between matter and photon correlations requires that excitons thermalize between successive emission events, i.e.\ $\tau_\text{th} \ll \Gamma^{-1}$, where $\tau_\text{th}$ is the exciton--phonon scattering time and $\Gamma^{-1}$ the radiative lifetime.

For monolayer TMD excitons at the temperatures of our experiment ($T \approx 3\text{--}7$ K), the dominant thermalization channel is coupling to acoustic phonons through the deformation-potential mechanism. For interlayer (vertically-stacked) excitons in MoSe$_2$/WSe$_2$ heterobilayers, the thermalization timescale has been measured directly by time-resolved photoemission~\cite{Policht2023}, yielding $\tau_\text{th} \sim 1\text{--}10$ ps.

The exciton--phonon coupling for the 1D interface excitons at the lateral MoSe$_2$/WSe$_2$ heterojunction studied here has not, to our knowledge, been calculated. However, the constituent electron and hole occupy the same monolayers as in the vertical interlayer-exciton case --- MoSe$_2$ for the electron and WSe$_2$ for the hole --- and therefore share essentially the same phonon environment and similar deformation-potential couplings. We thus adopt
\begin{equation}
    \tau_\text{th} \;\sim\; 1\text{--}10\,\text{ps}
    \tag{S1}
\end{equation}
as an order-of-magnitude estimate for our system. Combined with the independently measured radiative lifetime $\Gamma^{-1} \sim 5\text{--}15$ ns of the trapped 1D excitons~\cite{Vandoolaeghe2025}, the hierarchy
\begin{equation}
    \frac{\Gamma^{-1}}{\tau_\text{th}} \;\sim\; 10^3\text{--}10^4
    \tag{S2}
\end{equation}
implies that the matter undergoes thousands of thermalization cycles between successive emission events. The quasi-equilibrium reading of Eq.~(1) is therefore well-justified across the full power range of our experiments.

\subsection*{Hierarchy and validity}

Collecting (S4), (S7), and (S8):
\begin{equation}
    \tau_\text{int}\ (\sim 0.05\text{--}3\,\text{ps}) \;\ll\; \tau_\text{th}\ (\sim 1\text{--}10\,\text{ps}) \;\ll\; \Gamma^{-1}\ (\sim 5\text{--}15\,\text{ns}),
    \tag{S9}
\end{equation}
with $\Gamma^{-1}/\tau_\text{th} \sim 10^3\text{--}10^4$.  The matter thus undergoes $\sim\!10^3\text{--}10^4$ thermalization cycles between successive radiative emission events.  Each emitted photon samples a thermalized configuration of the gas, validating the quasi-equilibrium reading of Eq.~(1) in the main text. The hierarchy is preserved across the full power range: $\tau_\text{int}$ shortens with increasing density, and $\tau_\text{th}$ remains bounded above by the few-ps phonon scattering time, so condition (S1) holds uniformly across the bunching-to-antibunching crossover.

\subsection{Theoretical model for confined 1D dipolar excitons}
\subsubsection{Many-Body Framework}

The conventional theoretical approach to the physics of degenerate Bose gases is the Gross--Pitaevskii equation (GPE), where the many-body state is represented by a single macroscopically occupied wave-function~\cite{legget2001bose}. Despite the generality and predictive power of the GPE, it encounters limitations in low-dimensional and few-body systems, where enhanced quantum fluctuations do not allow formation of a condensate, weakening the conditions to employ a macroscopic wave-function~\cite{haegeman2017quantum}. From another perspective, 1D Bose gases at strong interactions develop fermionic statistics, the so-called Tonks--Girardeau limit~\cite{Kinoshita2004}. In this regime, a many-body description would require a number of single-particle orbitals equal to the number of particles $N$. In our system, crystallization and sub-Poissonian statistics occur precisely in the strongly interacting regime, where we expect the particle to occupy different (localized) single particle orbitals.

A convenient framework to resolve this difficulty is the Multiconfigurational Time-Dependent Hartree method for bosons (MCTDH-B)~\cite{MCTDHB}, in which the many-body wavefunction is expanded over all configurations of $N$ bosons in $M$ time-dependent orbitals,
\begin{equation}
|\Psi(t)\rangle = \sum_{\{n_j\}} A_{n_1,\ldots,n_M}(t)\,|n_1,\ldots,n_M; t\rangle,
\label{eq:ansatz-general}
\end{equation}
with both the coefficients $A_{n_1,\ldots,n_M}(t)$ and the orbitals $\phi_j(\mathbf{r},t)$ determined variationally. The equations of motion follow from the Dirac--Frenkel variational principle, $\langle\delta\Psi\,|\,i\hbar\,\partial_t - \hat{H}\,|\,\Psi\rangle = 0$, which yields a set of two coupled differential equations: one for the coefficient vector, and other for the orbitals. The method recovers Gross--Pitaevskii for $M=1$ and becomes \emph{in principle} exact as $M\to\infty$.
To simulate our system of dipolar excitons in lateral heterostructures, we
consider $N$ interacting bosons in 2D geometry,
confined by a external potential: a quantum well along the $x$
direction defines the interface along which the excitons are free to move,
and a strong harmonic confinement along $y$.
The system is governed by the many-body Hamiltonian
\begin{equation}
    \hat{H} = \sum_{i=1}^{N} \hat{h}(\mathbf{r}_i)
    \;+\; \sum_{i<j}^{N} W(\mathbf{r}_i,\mathbf{r}_j),
    \label{eq:hamiltonian}
\end{equation}
where the single-particle Hamiltonian
\begin{equation}
    \hat{h}(\mathbf{r}) = -\frac{\hbar^2}{2m}\nabla^2
    \;+\; V_{x}(x)
    \;+\; \tfrac{1}{2}\,m\,\omega_y^{2}\,y^{2}
    \label{eq:single-particle}
\end{equation}
contains the kinetic energy, with $m$ the exciton mass, and the external potential.
The excitons are coupled through a two-body interaction $W(\mathbf{r},\mathbf{r}')$,
specified in the next section.

We solve the many-body problem defined by Eq.~\eqref{eq:hamiltonian} using
the MCTDH ansatz Eq.~\eqref{eq:ansatz-general}. 
 Once $W$ is specified, the equations are completely defined and can be propagated numerically. So we need to determine the appropriate form of $W(X,Y)$ for our dipolar exciton system, which is the subject of the next section.
\subsubsection{Single-exciton Hamiltonian and numerical solution}
 The interaction between two dipolar excitons in a lateral type-II heterostructure is inherited from the internal structure of the exciton.
 % : its permanent dipole moment $d$ sets the long-range dipolar tail, and its finite size regularizes the short-range divergence. 
 Before computing the exciton--exciton interaction, we must therefore solve the single-exciton problem.
% \subsection{Effective-mass Hamiltonian}

We consider an electron--hole pair at a lateral type-II interface between
two monolayer transition-metal dichalcogenides (TMDs), with the interface
along the $x$ axis. In the effective-mass approximation~\cite{Lau2018}, the Hamiltonian for an exciton can be written as,
\begin{equation}
\begin{aligned}
H_{\mathrm{exc}} \;=\;
  &-\frac{\hbar^{2}}{2 m_e}\nabla_{\mathbf{r}_e}^{2}
   \;-\; \frac{\hbar^{2}}{2 m_h}\nabla_{\mathbf{r}_h}^{2} \\[4pt]
  &+\; V_{\mathrm{int}}(\mathbf{r}_e,\mathbf{r}_h)
   \;+\; V_e(\mathbf{r}_e) \;+\; V_h(\mathbf{r}_h),
\end{aligned}
\label{eq:H-exciton}
\end{equation}
where $m_e$ and $m_h$ are the electron and hole effective masses,
$\mathbf{r}_e$ and $\mathbf{r}_h$ their in-plane coordinates, and
$V_{\mathrm{int}}$ is the Coulomb attraction between the electron and the
hole. In the two-dimensional limit, this attraction is given by the
Rytova--Keldysh potential~\cite{Cudazzo2011,Berkelbach2013},
\begin{equation}
V_{\mathrm{int}}(r) \;=\;
  -\,\frac{e^{2}\pi}{2\, r_s}
  \left[\,
    \mathbf{H}_{0}\!\left(\frac{r}{r_s}\right)
    - \mathbf{Y}_{0}\!\left(\frac{r}{r_s}\right)
  \right],
\label{eq:RK}
\end{equation}
where $r=|\mathbf{r}_e-\mathbf{r}_h|$ is the electron--hole separation,
$e$ the elementary charge, $r_{s}$ the screening length set by the
2D polarizability of the TMD monolayers, and $\mathbf{H}_0$
and $\mathbf{Y}_0$ are, respectively, the Struve function and the Bessel
function of the second kind, both of order zero.

The type-II band offset across the lateral interface is modelled as a
smooth step,
\begin{equation}
V_{e}(y) \;=\; -V_0\,\tanh\!\left(\frac{y}{w}\right),
\qquad
V_{h}(y) \;=\; +V_0\,\tanh\!\left(\frac{y}{w}\right),
\label{eq:band-offset}
\end{equation}
where $V_0$ is the band offset and $w$ the interface width; the opposite
signs confine the electron and the hole to opposite sides of the
interface. Hence, the confinement potential can be defined as
\begin{equation}
V_{\mathrm{conf}}(\mathbf{r}_e,\mathbf{r}_h)
  \;=\; V_e(\mathbf{r}_e) \;+\; V_h(\mathbf{r}_h).
\label{eq:Vconf}
\end{equation}

As the longitudinal well is much more extended than the transverse trapping $L\gg \ell_{y}$, we assume translational invariance and conservation of the longitudinal component of the Centre-of-Mass (COM) momentum. Within this assumption, we rewrite the Hamiltonian in COM and relative coordinates. The Hamiltonian reduces to a three-coordinate problem in $(Y, r_x, r_y)$,
\begin{equation}
\hat{H}_{\mathrm{exc}} \;=\; -\frac{\hbar^{2}}{2M}\,\partial_{Y}^{2}
                       \;-\; \frac{\hbar^{2}}{2\mu}\partial_{\mathbf{r}^{2}}^{2}
                       \;+\; V_{\mathrm{int}}(\mathbf{r})
                       \;+\; V_{\mathrm{conf}}(Y,\mathbf{r}).
\label{eq:H-3coord}
\end{equation}
where $M=m_e+m_h$, and $\mu=m_e m_h/M$ the reduced mass,
$Y$ the transverse COM coordinate, and $\mathbf{r}=(r_x,r_y)$ the relative
electron--hole coordinate.
We solve Eq.~\eqref{eq:H-3coord} by exact diagonalization (ED) using the locally optimal block preconditioned conjugate gradient (LOBPCG) method to obtain the lowest eigenstate. For given experimental parameters, we
obtain the ground-state energy $E_0$ and the corresponding wavefunction
$\Psi(Y, r_x, r_y)$. From this wavefunction we compute the permanent dipole
length $d = \langle r_y \rangle$ and the spatial spread $\Delta r_y$.

\subsubsection{Effective exciton--exciton interaction}

The bare interaction between two excitons, with centres of mass at
$\mathbf{R}_1$, $\mathbf{R}_2$ and internal electron--hole coordinates
$\mathbf{r}_1$, $\mathbf{r}_2$, is the sum of the four pairwise Coulomb
terms between their electrons and holes. Grouping these into a single
interaction $V_{\mathrm{int}}(\mathbf{R}; \mathbf{r}_1, \mathbf{r}_2)$, the
net interaction between the excitons as composite particles is obtained by
averaging over their internal ground states,
\begin{equation}
W(\mathbf{R}) \;=\; \!\int\! d\mathbf{r}_1\, d\mathbf{r}_2 \;
                \rho_{\mathrm{rel}}(\mathbf{r}_1)\,
                \rho_{\mathrm{rel}}(\mathbf{r}_2)\;
                V_{\mathrm{int}}(\mathbf{R}; \mathbf{r}_1, \mathbf{r}_2),
\label{eq:W-formula}
\end{equation}
where $\mathbf{R} = \mathbf{R}_1 - \mathbf{R}_2$ is the COM separation and
$\rho_{\mathrm{rel}}(\mathbf{r}) = |\psi_{\mathrm{rel}}(\mathbf{r})|^{2}$ is
the internal density of a single exciton in its ground state.

To evaluate Eq.~\eqref{eq:W-formula}, we fit the relative-motion amplitude
$\psi_{\mathrm{eff}}(r_x, r_y) \equiv \sqrt{\rho_{\mathrm{rel}}(r_x, r_y)}$
from the ED ground state to a compact analytic ansatz,
\begin{equation}
\psi_{\mathrm{eff}}(r_x, r_y) \;=\;
  B \,\exp\!\left[ -\sqrt{\,\frac{r_x^{2}}{\lambda_x^{2}}
  + \frac{(r_y-d)^{2}}{\lambda_y^{2}}}\,\right],
\label{eq:psi-ansatz}
\end{equation}
an exponential decay capturing the asymptotic form of a Coulomb-bound
state, centred on the dipole displacement $d$ and made elliptical by the
rescaling $(\lambda_x, \lambda_y)$ along and transverse to the dipole axis;
$B$ is fixed by normalization. 

Substituting $\rho_{\mathrm{rel}} = |\psi_{\mathrm{eff}}|^{2}$ into Eq.~\eqref{eq:W-formula} yields the effective interaction $W(X, Y)$, which takes the form of a regularized dipole--dipole potential with a $1/R^{3}$ tail whose strength is set by the dipole length $d$. By construction, $W(X, Y)$ is attractive for head-to-tail exciton configurations. In the experimental geometry, however, such configurations are strongly suppressed and the two excitons interact predominantly in a side-to-side arrangement. Consequently, although the effective potential develops attractive lobes, their thermodynamic weight is negligible.

Nevertheless, these attractive lobes give rise to prohibitively long thermalization times in the MCTDHB simulations, likely because their strength is overestimated within our  derivation. To circumvent this numerical difficulty, we remove the attractive lobes from the interaction potential and employ a repulsive-only fit throughout.
\begin{equation}
W_{\mathrm{fit}}(X, Y) \;=\;
  \frac{C_{dd}}{\left(\,X^{2} + c^{2} Y^{2}
  + \varepsilon^{2}\,\right)^{3/2}},
\label{eq:W-fit}
\end{equation}
where $C_{dd}$ is the dipolar coupling strength, $c$ the in-plane
anisotropy, and $\varepsilon$ the short-range regularization length set by
the exciton's spatial extent. This provides the closed-form interaction
used in the MCTDHB Hamiltonian, Eq.~\eqref{eq:hamiltonian}.
\begin{figure}[!t]
    \centering
    \includegraphics[width=\columnwidth]{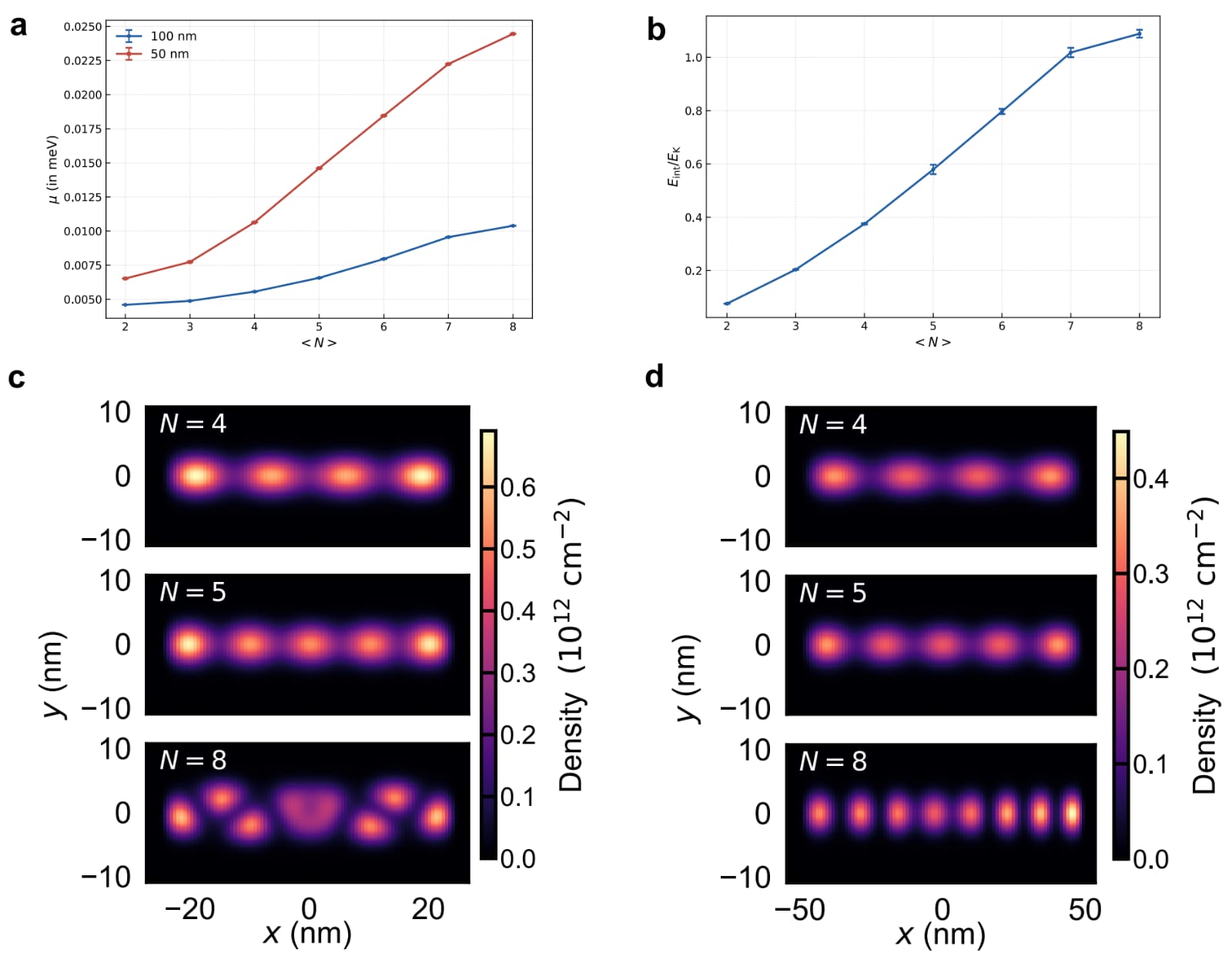}
    \caption{\textbf{Many-body ground state of dipolar excitons across the
    interaction crossover.}
    \textbf{(a)} Chemical potential $\mu = \partial E/\partial N$ versus
    $N$ for both 50nm and 100nm traps;
    error bars reflect the DVR-truncation uncertainty.
    \textbf{(b)} Interaction-to-kinetic energy ratio
    $\gamma = E_{\mathrm{int}}/E_{\mathrm{kin}}$, which scales linearly as a function of $N$. The
    crossover from $\gamma \lesssim 1$ to $\gamma \gg 1$ marks the transition from the weakly correlated to the strongly correlated
    regime.
    \textbf{(c,d)} Computed $T = 0$ density distributions
    $n(x) = \langle\hat{\Psi}^{\dagger}(x)\hat{\Psi}(x)\rangle$ for the
    $50\,$nm trap (c) and the $100\,$nm trap (d), shown for
    $N = 4, 5, 8$ (top to bottom). In the tightly confined
    $50\,$nm trap the $N = 8$ state develops a staggered, zig-zag
    arrangement of the density peaks, signalling the onset of transverse
    ordering; no such staggering is resolved in the $100\,$nm trap, where
    the weaker confinement keeps the density modulation purely
    longitudinal. The $N = 8$ calculations require substantially larger
    orbital and DVR bases, and within our computational budget
    convergence could be reached only to an energy tolerance of
    $\Delta E \approx 10^{-5}$.}
    \label{fig:supp_theory}
 
\end{figure}
\subsubsection{Results}
We perform many-body simulations of dipolar excitons for two trap lengths,
$L = 50\,$nm and $L = 100\,$nm. For each trap, we propagate the MCTDHB
equations of motion in imaginary time until the energy converges to
$\Delta E < 10^{-7}$, yielding the interacting ground state. We consider
particle numbers $N = 2$--$8$. Convergence with respect to the number of
orbitals $M$ was checked individually: for $N = 2$ and $N = 3$ we used
$M = 3$ and $M = 8$, respectively, with the higher orbitals carrying
negligible population; for $N \geq 4$ we used $M = 10$. In the crystalline
phase at high density, the occupation of high-energy orbitals remains
relatively large, increasing the computational cost. The choice $M = 10$
represents a good compromise between accuracy and efficiency.

From the converged ground-state energies $E(N)$ we extract the chemical
potential as a discrete derivative,
\begin{equation}
\mu(N) \;=\; \frac{\partial E}{\partial N}
        \;\approx\; E(N) - E(N-1),
\label{eq:mu}
\end{equation}
and inverse compressibility from the second derivative,
\begin{equation}
\kappa^{-1} \;\propto\; \frac{L}{N}\,\frac{\partial^{2} E}{\partial N^{2}},
\label{eq:kappa}
\end{equation}
which quantifies the energy cost of compressing the dipolar exciton gas.
The chemical potential extracted from Eq.~\eqref{eq:mu} is shown in
Fig.~\ref{fig:supp_theory}(a) for both trap lengths. It grows with $N$
and tends to saturate once the dipoles begin to explore the transverse
direction, as accommodating additional particles at high density requires
populating higher transverse modes. To characterise the interaction
crossover, we also compute the interaction-to-kinetic energy ratio
$E_{\mathrm{int}}/E_{\mathrm{kin}}$, plotted in
Fig.~\ref{fig:supp_theory}(b). This ratio scales linearly with $N$ and
marks the transition from the weakly to the strongly correlated regime.

The corresponding $T = 0$ density distributions
$n(x) = \langle\hat{\Psi}^{\dagger}(x)\hat{\Psi}(x)\rangle$ are displayed
in Fig.~\ref{fig:supp_theory}(c,d) for particle numbers
$N = 4, 5, 8$. In the tightly confined $50\,$nm trap
[Fig.~\ref{fig:supp_theory}(c)], the $N = 8$ state develops a zig-zag,
staggered arrangement of the density peaks, signalling the onset of
transverse ordering. No such staggering is observed in the $100\,$nm trap
[Fig.~\ref{fig:supp_theory}(d)], where the weaker confinement produces
density modulations only along the longitudinal direction.

Both ground-state relaxations are performed with the \texttt{MCTDH-X} package~\cite{Molignini2025MCTDHX,Lin2020QST} using the Davidson (DAV) algorithm for imaginary-time propagation. We have extended and modified the publicly available \texttt{MCTDH-X} code~\cite{mctdhx-releases-2025} to treat angle-dependent dipolar interactions.

\end{document}